\documentclass[10pt, conference, letterpaper]{IEEEtran}

\usepackage{times}
\usepackage{latexsym}
\usepackage{hyperref}
\usepackage{xcolor}
\usepackage{cite}
\usepackage{amsmath,amssymb,amsfonts}
\usepackage{stackrel}
\usepackage{algorithmic}
\usepackage{graphicx}
\usepackage{textcomp}

\newtheorem{theorem}{Theorem}

\newtheorem{condition}{Condition}

\title{Utility Optimal Scheduling with a Slow Time-Scale Index-Bias
  for Achieving Rate Guarantees in Cellular Networks}

\author{
\IEEEauthorblockN{Anurag Kumar}
\IEEEauthorblockA{\textit{ECE Department} \\
\textit{Indian Institute of Science}\\
Bengaluru, India \\
anurag@iisc.ac.in}
\and
\IEEEauthorblockN{Rajesh Sundaresan}
\IEEEauthorblockA{\textit{ECE Department} \\
\textit{Indian Institute of Science}\\
Bengaluru, India \\
rajeshs@iisc.ac.in}
}

\begin{document}
\maketitle

\begin{abstract}
  One of the requirements of \emph{network slicing} in 5G networks is
  RAN (radio access network) scheduling \emph{with rate guarantees.}
  We study a three-time-scale algorithm for maximum sum utility
  scheduling, with minimum rate constraints. As usual, the scheduler
  computes an index for each UE in each slot, and schedules the UE
  with the maximum index. This is at the fastest, natural time-scale
  of channel fading. The next time-scale is of the exponentially
  weighted moving average (EWMA) rate update. The slowest time scale
  in our algorithm is an ``index-bias'' update by a stochastic
  approximation algorithm, with a step-size smaller than the EWMA. The
  index-biases are related to Lagrange multipliers, and bias the slot
  indices of the UEs with rate guarantees, promoting their more
  frequent scheduling.  We obtain a pair of coupled ordinary
  differential equations (o.d.e.) such that the unique stable points
  of the two o.d.e.s are the primal and dual solutions of the
  constrained utility optimization problem. The UE rate and index-bias
  iterations track the asymptotic behaviour of the o.d.e. system for
  small step-sizes of the two slower time-scale iterations.
  Simulations show that, by running the index-bias iteration at a
  slower time-scale than the EWMA iteration and using the EWMA
  throughput itself in the index-bias update, the UE rates stabilize
  close to the optimum operating point on the rate region boundary,
  and the index-biases have small fluctuations around the optimum
  Lagrange multipliers. We compare our results with a prior 
  two-time-scale algorithm
  %\cite{winet.mandelli-andrews-borst-klein2019satisfying-network-slicing-constraints-via-5G-mac-scheduling}.
  and show improved performance.
\end{abstract}

\section{Introduction}
\label{sec:introduction}

An important functionality that has been planned for 5G cellular
networks, and beyond, is that of \emph{network slicing} (see, for
example,
\cite{winet.banchs-deveciana-etal2020resource-allocation-for-network-slicing-in-mobile-networks}). A
``slice'' in a 5G network would be a virtual subnetwork extending from
the point of Internet access, all the way through the transport and
core, up to and including the radio access network (RAN), configured
to provide some desired services to the user connections carried on
that slice. One such service could be that the connections carried by
the slice are guaranteed a minimum average aggregate throughput. With
the backhaul comprising high-speed optical network links, the
bottleneck in the provision of such rate guarantees would be the
RAN. In this paper, we are concerned with RAN schedulers that can also
provide rate guarantees.

We consider the downlink of a single base station (BS) with several
associated UEs (user equipments). The radio access is time slotted. In
each slot, the BS uses the information about the rates that can be
alloted to the UEs in the slot, along with the average throughputs
achieved thus far by the UEs, to dynamically determine the rates to be
allotted to the UEs in the current slot. The popular approach is based
on a dynamic allocation scheme that, in the long run, results in
average UE throughputs that maximize the total utility over the UEs,
where each UE's utility is a concave, increasing function of its
throughput (see, for example,
\cite{winet.kushner-whiting04convergence-proportional-fairness}). Network
slicing motivates sum utility optimal scheduling with minimum rate
guarantees. We consider the provision of individual minimum rate
guarantees to some of the UEs, assuming that the guarantees are
feasible, and provide an extension of the usual dynamic scheduling
algorithm such that resulting average throughputs respect the desired
minimum rate guarantees.

\subsection{Related Literature}
\label{sec:related-literature}

After the original proportional fair scheduling algorithm was
introduced (using log of throughput as the utility), there was a
series of papers that provided a deeper understanding, and proofs of
convergence of generalizations of the algorithm to other concave
nondecreasing utility functions; see
\cite{winet.agrawal-subramanian2002optimality-channel-aware-scheduling},
\cite{winet.kushner-whiting04convergence-proportional-fairness},
\cite{winet.stolyar05gradient-scheduling}; see also
\cite{winet.raman-jagannathan2018downlink-resource-allocation-under-time-varying-interference-fairness-and-throughput-optimality}
for a recent treatment of the problem. The approach has been to
consider a sequence of instances of the algorithm with reducing
weights for the EWMA (exponentially weighted moving average)
throughput calculation (i.e., increasing averaging window). It is
shown, then, by fluid limit arguments, that the EWMA throughputs track
a certain ordinary differential equation (o.d.e.), whose unique stable
point is the solution of the utility optimization over a rate region
that is the average of the slot-wise rate regions, the average being
taken over the stationary probability distribution of the joint
channel state Markov chain.

After the introduction of the network slicing concept in the 5G
cellular network standards, there has been interest in scheduling with
rate guarantees. An important precursor to our work reported in this
paper is the work by Mandelli et
al.~\cite{winet.mandelli-andrews-borst-klein2019satisfying-network-slicing-constraints-via-5G-mac-scheduling},
where the authors modify the calculation of certain indices used in
the classical scheduler by an additive \emph{bias}. This bias is
updated by an additive correction obtained by comparing the \emph{rate
  allocated in the current slot} against the minimum guaranteed rate.
The authors, essentially, extend an algorithm provided in
\cite[Section
6]{winet.stolyar2006greedy-primal-dual-algorithm-resource-allocation-complex-networks},
and utilize the convergence proof of
\cite{winet.stolyar05gradient-scheduling}. The EWMA update and the
bias update have a common time-scale. Simulation results are presented
that show that the network utility improves as compared to other
proposals for scheduling with rate guarantees.

\subsection{Contributions and Outline}
\label{sec:outline-contributions}

In our work, we have considered dynamic rate allocation algorithms for
the minimum rate constrained utility optimization problem, assuming
that the problem is feasible. It is easy to see
(Section~\ref{sec:motivation-for-algorithm}) that the index-biases
that promote the rate guarantees are the Lagrange multipliers
in the optimal solution of the utility maximization problem. We use a
stochastic approximation (SA) algorithm to update the Lagrange
multipliers \emph{based on the EWMA rates}, rather than the allocated
rates in a slot (see
Section~\ref{sec:gradient-scheduling-rate-guarantees}). The step size
in the SA update is smaller than the averaging weight used in the EWMA
calculation. Thus, in addition to the natural time-scale at which the
channel is varying slot-to-slot, there are two, successively slower,
time-scales, namely, the EWMA update time-scale for the throughputs and the
SA update time-scale for the Lagrange multipliers.

In comparison with the algorithm proposed in
\cite{winet.mandelli-andrews-borst-klein2019satisfying-network-slicing-constraints-via-5G-mac-scheduling},
we find that our algorithm provides (i) accurate convergence to the
desired operating throughputs with rate guarantees, and (ii) smooth
behaviour of the index-biases which are approximations to the Lagrange
multipliers. The two step-sizes (of the two SA iterations) can
predictably control the rate of convergence and accuracy (see
Section~\ref{sec:simulation-experiments}). Accurate estimates of the
Lagrange multipliers are useful to the network operator for the
purposes of pricing and admission control. Further, the standard or
user requirements may specify the throughput averaging time-scale, in
which case, a separate time-scale for the index-bias update gives the
operator the flexibility to ensure that the Lagrange multipliers are
estimated accurately. See, for example, \cite[Table
5.7.4-1]{winet.3GPP-5G-NR_system-specifications_etsi_138214v160200p}
where the averaging window for guaranteed rate services is given as
$2000$~ms.

In
Section~\ref{sec:convergence-analysis-gradient-scheduling-rate-guarantees},
under certain conditions, we show that the UE throughputs and the
index-biases track the trajectories of two coupled o.d.e.s, whose
unique stable point is the optimal primal-dual solution of the minimum
rate constrained utility optimization problem.

\section{System Model, Notation, \& Problem Definition}
\label{sec:system-model-and-notation}

There is a single gNB (generically a Base Station (BS)). The slots are
indexed by $k, k \geq 0$. Vectors are typeset in bold font, and are
viewed as column vectors.

\begin{description}

\item[$M$:] The number of UEs, indexed by $i, 1 \leq i \leq M$

\item[$\mathcal{S}$:] The set of joint channel states for the UEs; in
  a slot, the state $s \in \mathcal{S}$ is known to the scheduler, and
  fixes the number of bits each UE can transmit in the slot

\item[$S(k)$:] For $k \geq 0$, is the joint channel state process,
  assumed to be a Markov chain on $\mathcal{S}$, whose realization
  $s(k)$ in a slot~$k$ \emph{is known to the scheduler}

\item[$P(\cdot|s)$:] Transition probability kernel of the
  channel state process Markov chain $\{S_k, k \geq 0\}$

\item[$\pi_s$:] ($s \in \mathcal{S}$) is the stationary probability
  of the channel state process $\{S_k, k \geq 0\}$

\item[$\mathcal{R}_s$:] $\subset \mathbb{R}^M_+$; the joint rate
  region for the $M$ UEs in a slot when the channel state in the slot
  is $s \in \mathcal{S}$; for every $s \in \mathcal{S}$,
  $\mathcal{R}_s$ is convex and coordinate convex\footnote{A set is
    coordinate convex if $\mathbf{r} \in \mathcal{R}_s$ implies
    $\{\mathbf{x}: \mathbf{0} \leq \mathbf{x} \leq \mathbf{r}\}
    \subset \mathcal{R}_s$.}

\item[$\overline{\mathcal{R}}_{\cdot|s}$:] $\subset \mathbb{R}^M_+$;
  the rate region of the vector of average rates over a slot,
  given that the previous state is $s$; clearly,
  $\overline{\mathcal{R}}_{\cdot|s} = \sum_{s' \in \mathcal{S}} P(s' |
  s) \mathcal{R}_{s'}$; by the assumptions on $\mathcal{R}_{s'}$,
  $\overline{\mathcal{R}}_{\cdot|s}$ is convex and coordinate convex

\item[$\overline{\mathcal{R}}$:] $\subset \mathbb{R}^M_+$; the rate
  region of the vector of average rates over the slots; clearly,
  $\overline{\mathcal{R}} = \sum_{s \in \mathcal{S}} \pi_s
  \mathcal{R}_s$, and $\overline{\mathcal{R}}$ is convex and coordinate
  convex

  \item[$\hat{\mathcal{R}}$:]
  $([0, \hat{r}] \times [0, \hat{r}] \times \cdots \times [0,
  \hat{r}]) \subset \mathbb{R}^M_+$ where $\hat{r}$ is the maximum
  possible bit rate at which a gNB can transmit to a UE;
  $\mathcal{R}_s$, $\overline{\mathcal{R}}_{\cdot|s}$, $\overline{\mathcal{R}}$ are all subsets of
  $\hat{\mathcal{R}}$

\item[$U(\mathbf{r})$:] The utility of the rate vector
  $\mathbf{r} \in \mathcal{R}$; $U(\mathbf{r})$ is strictly concave,
  coordinate-wise nondecreasing over $\overline{\mathcal{R}}$, and
  continuously differentiable; typically,
  $U(\mathbf{r}) = \sum_{i=1}^M U_i(r_i)$, with $U_i(\cdot)$ strictly
  concave, nondecreasing, and continuously differentiable

\item[$\theta_i(k)$:] The EWMA throughput of $\text{UE}_i$ at the
  beginning of slot~$k$;
  $\boldsymbol{\theta}(k) = (\theta_1(k), \theta_2(k), \cdots,
  \theta_M(k))^T$, the column vector of all the EWMA throughputs at
  the beginning of slot~$k$

\item[$r_i(k)$:] The rate obtained by $\text{UE}_i$ in slot~$k$;
  $\mathbf{r}(k) = (r_1(k), r_2(k), \cdots, r_M(k))^T$, the column vector
  of rates obtained by all the UEs in slot~$k$

\item[$\theta_{i,\text{min}}$:] the rate guarantee for $\text{UE}_i$,
  possibly equal to $0$; ${\boldsymbol{\theta}}_{\text{min}}$ the
  vector of rate guarantees

\item[$\nu_i(k)$:] the nonnegative \emph{index bias} used for
  enforcing the rate guarantee for $\text{UE}_i$
  $\boldsymbol{\nu}(k) = (\nu_1(k), \nu_2(k), \cdots, \nu_M(k))^T$

\end{description}

\noindent
The {\em optimization problem} we consider in this work is:
\begin{eqnarray}
  \label{eqn:optimization-problem}
  \max U(\mathbf{r}) \text{ subject to }
  \mathbf{r} \in \overline{\mathcal{R}} \cap \{\mathbf{r} \in \mathbb{R}_+^M: \mathbf{r} \geq \boldsymbol{\theta}_{\text{min}}\}
\end{eqnarray}
Assuming that Problem~(\ref{eqn:optimization-problem}) is feasible,
due to the strict concavity of $U(\cdot)$ and the convexity of the
rate region $\overline{\mathcal{R}}$, we have a unique vector of optimum rates
$\boldsymbol{\theta}^* \in \overline{\mathcal{R}}$. \\

\noindent
The {\em scheduling problem} is: in a slot, given the state $s(k)$ and the
EWMA throughputs $\boldsymbol{\theta}(k)$, determine a rate assignment
$\mathbf{r}(k)$ so that the vector of EWMA rates eventually
gets close (in some sense) to $\boldsymbol{\theta}^*$.

\section{Scheduling with Rate Guarantees}
\label{sec:algorithm-pf-rg}

\subsection{Motivation for the algorithm}
\label{sec:motivation-for-algorithm}

Denoting by $co\{\cdot\}$ the convex hull of a set, consider the
special case where, for each $s \in \mathcal{S}$,
$\mathcal{R}_s = co\{\mathbf{r}_{s,i}, 1 \leq i \leq M\}$, where
$\mathbf{r}_{s,i} = (0, \cdots, 0, r_{s,i}, 0, \cdots, 0)$, with
$r_{s,i} \geq 0$, at the $i^{th}$ position, the rate $\text{UE}_i$
gets if scheduled alone when the channel state is $s$. This special
case corresponds to the situation in which, in each slot, any one UE
is scheduled.

For a scheduling algorithm, for an $s \in \mathcal{S}$ and
$1 \leq i \leq M$, let $x_{s,i}$ be the fraction of slots in which
$\text{UE}_i$ is scheduled when the channel state is $s$. Here, for an
$s \in \mathcal{S}$, $x_{s,i} \geq 0, 1 \leq i \leq M$, and
$\sum_{i=1}^M x_{s,i} \leq 1$. Clearly, any
$\mathbf{r} \in \mathcal{R}_s$ can be achieved by such time-sharing
over slots when the channel state is $s$.

The optimization problem (\ref{eqn:optimization-problem}) becomes
\begin{eqnarray}
  \label{eqn:optimisation-problem-scheduling-fractions}
  \max &\sum_{i=1}^M& U_i(\theta_i) \nonumber \\
  \theta_i &=& \sum_{s \in \mathcal{S}} \pi_s x_{s,i} r_{s,i} \nonumber \\
  \theta_i &\geq& \theta_{i,\text{min}} \nonumber \\
   \text{for all} \  s \in \mathcal{S}, \sum_{i=1}^M x_{s,i} \leq 1, &\text{and}& x_{s,i} \geq 0
\end{eqnarray}

Since the $U_i(\cdot)$ are strictly concave, we are optimizing a
concave function over linear constraints. Hence, the
Karush-Kuhn-Tucker (KKT) conditions are necessary and sufficient.

For $s \in \mathcal{S}$ and $1 \leq i \leq M$, $\nu_i \geq 0$, $\lambda_s \geq 0$ and
$\eta_{s,i} \geq 0$, we have the Lagrangian function
\begin{eqnarray*}
  \lefteqn{L(\mathbf{x}, \boldsymbol{\nu}, \boldsymbol{\lambda}, \boldsymbol{\eta})  :=} \\
   &&  \sum_{i=1}^M \left(U_i(\sum_{s \in \mathcal{S}} \pi_s x_{s,i} r_{s,i} )
    + \nu_i (\sum_{s \in \mathcal{S}} \pi_s x_{s,i} r_{s,i} - \theta_{i,\text{min}}) \right) \\
 &&+ \sum_{s \in \mathcal{S}} \lambda_s \left(1 - \sum_{i=1}^M x_{s,i} \right) + \sum_{s \in \mathcal{S}}\sum_{i=1}^M \eta_{s,i} x_{s,i}
\end{eqnarray*}
We need a KKT point
$(\mathbf{x}^*, \boldsymbol{\nu}^*, \boldsymbol{\lambda}^*, \boldsymbol{\eta}^*)$ that
satisfies (writing
$\theta_i^* = \sum_{s \in \mathcal{S}} \pi_s x_{s,i}^* r_{s,i}$), for
every $s \in \mathcal{S}$ and $1 \leq i \leq M$:
\begin{eqnarray}
  \label{eqn:kkt-conditions}
  \frac{d}{d \theta_i}U_i(\theta_i^*) \pi_s r_{s,i} + \nu_i^* \pi_s r_{s,i} - \lambda_s^* + \eta_{s,i}^* &=& 0 \nonumber \\
  \nu_i^*(\theta_i^* - \theta_{i,\text{min}}) &=& 0 \nonumber \\
  \lambda_s^* (1 - \sum_{i=1}^M x_{s,i}^*) &=& 0 \nonumber \\
  \eta_{s,i}^* x_{s,i}^* &=& 0
\end{eqnarray}
Rewriting the first of these equations, we get
\begin{eqnarray*}
  \left(\frac{d}{d \theta_i}U_i(\theta_i^*)  + \nu_i^*\right)  r_{s,i} &=& \frac{1}{\pi_s} (\lambda_s^* - \eta_{s,i}^*)
\end{eqnarray*}
From the fourth condition (a complementary slackness condition) among
those in Equation~(\ref{eqn:kkt-conditions}), we see that whenever
$\eta_{s,i}^* > 0$ it must hold that $x_{s,i}^* = 0$, hence, for the
$(s,i)$ such that $x_{s,i} > 0$ the following must be satisfied
\begin{eqnarray*}
   \left(\frac{d}{d \theta_i}U_i(\theta_i^*)  + \nu_i^*\right)  r_{s,i} &=& \frac{1}{\pi_s} \lambda_s^*
\end{eqnarray*}
i.e., that
$\left(\frac{d}{d \theta_i}U_i(\theta_i^*) + \nu_i^*\right) r_{s,i}$
must be equalized across users $i$. It is this that motivates
Equation~(\ref{eqn:scheduled-rate-vector}) in our algorithm.

Note that, without the rate guarantee constraints, we get back the
original fair scheduling algorithm. The above analysis suggests that
the way forward is to incorporate the rate guarantees via the
corresponding Lagrange multipliers. Due to the additive manner in
which the $\nu_{i}$ appear, we will call them \emph{index biases}, as
they \emph{bias} the usual fair scheduling indices to encourage more
frequent scheduling of the UEs that require rate guarantees.

\subsection{Utility optimal scheduling with rate guarantees}
\label{sec:gradient-scheduling-rate-guarantees}

Two small step-sizes $a > 0, b > 0$, with $b \ll a$ are chosen. In
slot~$k$, given the EWMA throughputs $\boldsymbol{\theta}^{(a,b)}(k)$,
the index biases, $\nu_i^{(a,b)}(k) \geq 0, 1 \leq i \leq M$, and the
joint channel state $s(k) \in \mathcal{S}$, determine
$\mathbf{r}^*(\boldsymbol{\theta}^{(a,b)}(k),
\boldsymbol{\nu}^{(a,b)}(k), s(k))$ as follows:
\begin{eqnarray}
  \label{eqn:scheduled-rate-vector}
 \lefteqn{ \mathbf{r}^*(\boldsymbol{\theta}^{(a,b)}(k), \boldsymbol{\nu}^{(a,b)}(k), s(k)) =} \nonumber \\
  && \arg \max_{\mathbf{r} \in \mathcal{R}_{s(k)}} (\nabla U(\boldsymbol{\theta}^{(a,b)}(k)) +
  \boldsymbol{\nu}^{(a,b)}(k))^T\cdot\mathbf{r}
\end{eqnarray}
Update the EWMA throughputs as follows:
\begin{eqnarray}
  \label{eqn:ewma-rate-update}
 \lefteqn{ \boldsymbol{\theta}^{(a,b)}(k+1) = \boldsymbol{\theta}^{(a,b)}(k)} \nonumber \\
 && + a(\mathbf{r}^*(\boldsymbol{\theta}^{(a,b)}(k), \boldsymbol{\nu}^{(a,b)}(k), s(k)) -
  \boldsymbol{\theta}^{(a,b)}(k))
\end{eqnarray}
For all $i$, update the index biases as follows:
\begin{eqnarray}
  \label{eqn:index-bias-update}
  \boldsymbol{\nu}^{(a,b)}(k+1) = \left[\boldsymbol{\nu}^{(a,b)}(k) + b(\boldsymbol{\theta}^{(a,b)}_{\text{min}} -
  \boldsymbol{\theta}^{(a,b)}(k))\right]_0^{\nu_{\text{max}}}
\end{eqnarray}
$\nu_{\text{max}}$ is a suitably chosen value that is large enough so
that the equilibrium values of the index-biases lie within
$[0, \nu_{\text{max}})^M$. We will discuss a condition for ensuring
this in Section~\ref{sec:stable-points-and-the-optimum-solution}

 %Swap the last two sentences to bring the changes to nu/tau updates together. Then speak about how it is used to bias the throughput iterate.
The above algorithm needs to be compared with a similar one provided
in
\cite{winet.mandelli-andrews-borst-klein2019satisfying-network-slicing-constraints-via-5G-mac-scheduling}. The
difference lies in the recursion for the index bias,
Equation~\eqref{eqn:index-bias-update}. The recursion above is of the
stochastic approximation type, whereas the one in
\cite{winet.mandelli-andrews-borst-klein2019satisfying-network-slicing-constraints-via-5G-mac-scheduling}
is similar to the recursion for a queue (called a \emph{token counter}
in that paper). Another difference is that in
Equation~\eqref{eqn:index-bias-update} we use the EWMA throughput in
the index-bias update, whereas in
\cite{winet.mandelli-andrews-borst-klein2019satisfying-network-slicing-constraints-via-5G-mac-scheduling}
just the rate allocated in the current slot is used; see
Section~\ref{sec:pf-rg-tc-algorithm}.In
\cite{winet.mandelli-andrews-borst-klein2019satisfying-network-slicing-constraints-via-5G-mac-scheduling},
the token counter is used in
Equation~(\ref{eqn:scheduled-rate-vector}) after multiplication by the
same step-size, $a$, as is used in the EWMA throughput calculation.

\section{Convergence Analysis}
\label{sec:convergence-analysis-gradient-scheduling-rate-guarantees}

In Section~\ref{sec:convergence-analysis-with-rg-two-ues}, to provide
insight, we give a rough sketch of the proof for two UEs. The rigorous
proof of the formal convergence statement is provided in
Section~\ref{sec:weak-convergence-with-rate-guarantees}.

\subsection{Outline of the convergence analysis: a two UE example}
\label{sec:convergence-analysis-with-rg-two-ues}

There are 2 UEs with $\theta_{0,\text{min}} = 0$ and
$\theta_{1,\text{min}} >0$. Hence, we only need to update the index
bias $\nu_1(k), k \geq 0$. We have to analyze the following recursions:
\begin{eqnarray}
  \label{eqn:recursions-with-rg-two-ues}
  \lefteqn{\mathbf{r}(k) =} \nonumber \\
&& \hspace{-25mm} \arg \max_{\mathbf{r} \in \mathcal{R}_{s(k)}}
  \left[\frac{d}{d\theta_0} U(\theta_0(k)) \
                    (\frac{d}{d\theta_1} U(\theta_1(k)) + {\nu}_1(k))\right]\cdot\mathbf{r}  \\
  \boldsymbol{\theta}(k+1) &=& \boldsymbol{\theta}(k) + a(\mathbf{r}(k) - \boldsymbol{\theta}(k)) \\
  \nu_1(k+1) &=& (\nu_1(k) + b (\theta_{1,\text{min}} - \theta_1(k)))^+
\end{eqnarray}
Notice that if $\theta_1(k+1) < \theta_{1,\text{min}}$, the index bias
$\nu_1(k+1)$ increases, thereby providing a larger bias in favour of
$\text{UE}_1$. This encourages the $\text{UE}_1$ throughput to go towards $\theta_{1,\text{min}}$.

Since $b \ll a$, the index bias evolves slowly as compared to the EWMA
throughput updates.  Motivated by \cite[Chapter
6]{books.borkar2008stochastic-approximation_a-dynamical-systems-viewpoint} and \cite[Section
8.6]{books.kushner-yin2003stochastic-approximation-algorithms-with-applications},
we consider the index bias to be fixed at $\nu_1$, and define
\begin{eqnarray*}
  \mathbf{r}^*(\boldsymbol{\theta}, \nu_1, s) &=&
  \arg \max_{\mathbf{r} \in \mathcal{R}_{s}}
  \left[\frac{d}{d\theta_0} U(\theta_0) \
                    (\frac{d}{d\theta_1} U(\theta_1) + \nu_1)\right]\cdot\mathbf{r}
\end{eqnarray*}
Since $a$ is also small, in large time, the following
o.d.e.\ can be considered to hold:
\begin{eqnarray}
  \label{eqn:ode-with-rg-two-ues}
\dot{\boldsymbol{\theta}}(t) &=&
        \sum_{s \in \mathcal{S}} \pi_s \mathbf{r}^*(\boldsymbol{\theta}(t), \nu_1, s)  - \boldsymbol{\theta}(t)
\end{eqnarray}
Denote the stationary point of this o.d.e.\ by
$(\theta_{0,\infty}(\nu_1) \ \ \ \theta_{1,\infty}(\nu_1))$.  Since
$b \ll a$, the EWMA updates can be taken to have stabilized with
respect to the slower index-bias updates. With this in mind, the index
bias recursion in Equation~(\ref{eqn:recursions-with-rg-two-ues})
yields the following o.d.e.\ for the index bias for $\text{UE}_1$ (if
we assume that $\nu_1(t) > 0$ for all $t$):
\begin{eqnarray}
  \label{eqn:ode-index-bias-ue1}
  \dot{\nu}_1(t) = \theta_{1,\text{min}} - \theta_{1, \infty}(\nu_1(t))
\end{eqnarray}
The stationary point for this o.d.e.\ will yield
\begin{eqnarray}
  \label{eqn:stationary-point-index-bias-ode-ue1}
  \theta_{1,\infty} (\nu_{1, \infty}) &=& \theta_{1,\text{min}}
\end{eqnarray}
i.e., the limiting $\text{UE}_1$ throughput at the limiting index bias
is $\theta_{1,\text{min}}$.

We can infer that the following holds for the stationary point of the
o.d.e.\ in Equation~(\ref{eqn:ode-with-rg-two-ues}) with
$\nu_1 = \nu_{1,\infty}$:
\begin{eqnarray}
  \label{eqn:throughput-limits-in-terms-of-state-dependent-rate-region}
\lefteqn{\left[\theta_{0,\infty}(\nu_{1,\infty}) \  \theta_{1,\infty}(\nu_{1,\infty})\right] =
 \sum_{s \in \mathcal{S}} \pi_s  \arg \max_{\mathbf{r} \in \mathcal{R}_{s}} } \nonumber \\
  &&\hspace{-13mm} \left[\frac{d}{d\theta_0} U(\theta_{0,\infty}(\nu_{1,\infty})) \
    (\frac{d}{d\theta_1} U(\theta_{1,\infty}(\nu_{1, \infty})) + (\nu_{1, \infty}))\right]\cdot\mathbf{r}
\end{eqnarray}
It is easily seen that (see
\cite{winet.agrawal-subramanian2002optimality-channel-aware-scheduling},
\cite{winet.stolyar05gradient-scheduling}) under strict convexity of
$\overline{\mathcal{R}}$ (see
Condition~\ref{cond:strict-convexity-rate-regions}), for any
$\mathbf{x} \in \mathbf{R}^M_+$,
\begin{eqnarray}
  \label{eqn:equivalence-of-average-state-optimum-and-global-optimum}
  \arg \max_{\mathbf{r} \in \overline{\mathcal{R}}} \mathbf{x}^T\cdot\mathbf{r} &=&
   \sum_{s \in \mathcal{S}} \pi_s  \arg \max_{\mathbf{r} \in \mathcal{R}_{s}} \mathbf{x}^T\cdot\mathbf{r}
\end{eqnarray}
Using Equation~(\ref{eqn:equivalence-of-average-state-optimum-and-global-optimum}), we conclude that
\begin{eqnarray}
  \label{eqn:throughput-limits-in-terms-of-global-rate-region}
\lefteqn{\left[\theta_{0,\infty}(\nu_{1,\infty}) \ \ \ \theta_{1,\infty}(\nu_{1,\infty})\right] =
    \arg \max_{\mathbf{r} \in \overline{\mathcal{R}}} } \nonumber \\
 && \hspace{-13mm}   \left[\frac{d}{d\theta_0} U(\theta_{0,\infty}(\nu_{1,\infty})) \
    (\frac{d}{d\theta_1} U(\theta_{1,\infty}(\nu_{1, \infty})) + (\nu_{1, \infty}))\right]\cdot\mathbf{r}
\end{eqnarray}

\begin{figure}[h]
\begin{center}
\includegraphics[scale=0.4]{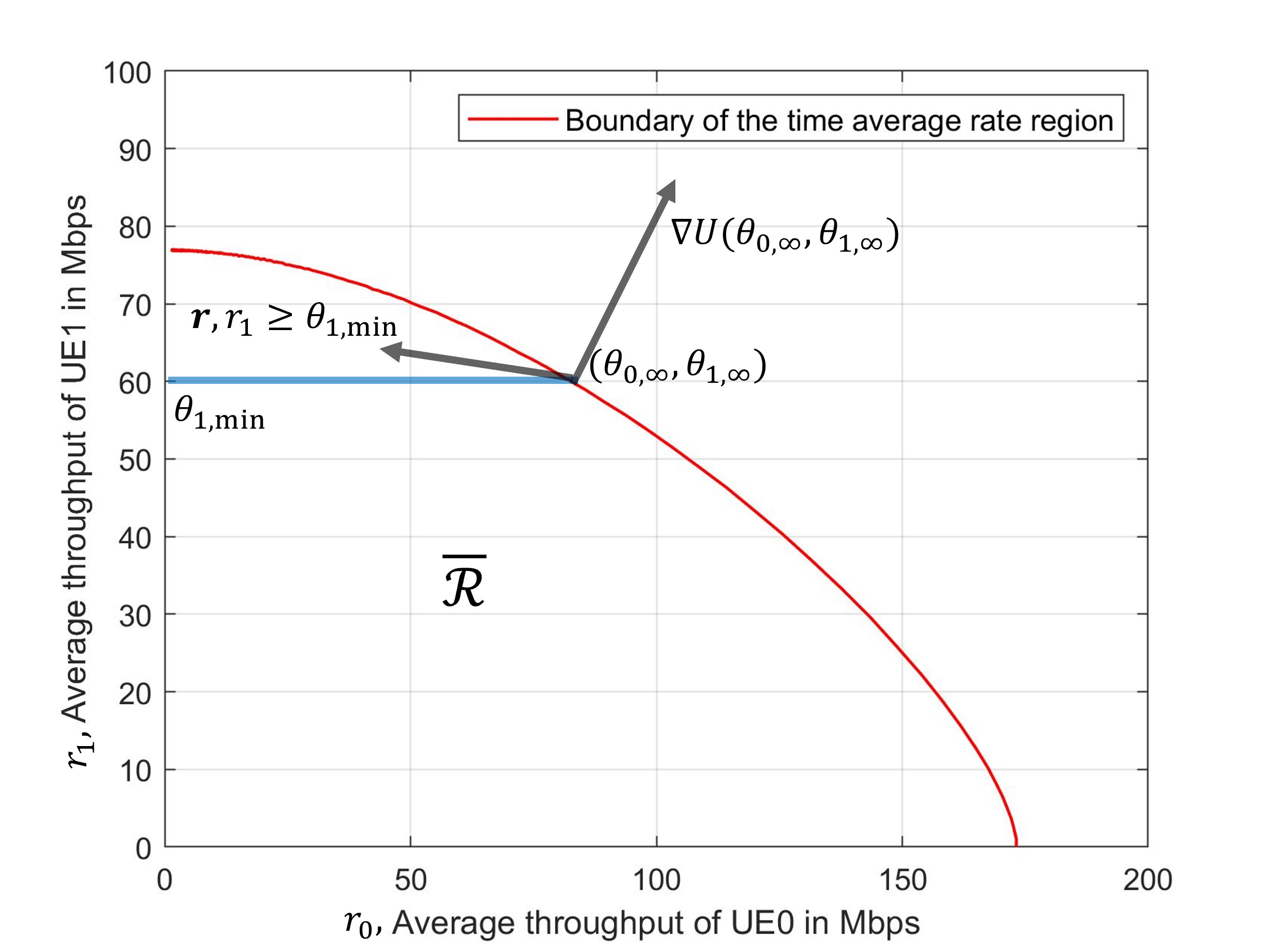}
\end{center}
\caption{Demonstration of the optimality of the o.d.e.\ limit. Two UEs
  at distances of $100$~m and $200$~m, system bandwidth $40$~MHz,
  slot time $1$~ms, gNB transmit power $100$~mW.}
\label{fig:showing-the-optimality-of-the-ode-limit}
\end{figure}

For notational convenience, let us write
$[\delta_0 \ \ \ \delta_1] := \left[\frac{d}{d\theta_0}
  U(\theta_{0,\infty}(\nu_{1,\infty})) \ \ \ \ (\frac{d}{d\theta_1}
  U(\theta_{1,\infty}(\nu_{1, \infty})) )\right]$. Then,
Equation~(\ref{eqn:throughput-limits-in-terms-of-global-rate-region}),
implies that, for all $\mathbf{r} \in \overline{\mathcal{R}}$
\begin{eqnarray}
  \label{eqn:establishing-optimality-with-rg-two-ues}
\lefteqn{  \left[\delta_0 \ \ \ \delta_1 + \nu_{1,\infty}\right].
  \left[\theta_{0,\infty}(\nu_{1,\infty}) \ \ \ \theta_{1,\infty}(\nu_{1,\infty})\right]^T \geq } \nonumber \\
 &&  \left[\delta_0 \ \ \ \delta_1+ \nu_{1,\infty}\right]\cdot\mathbf{r} \nonumber \\
\lefteqn{  \left[\delta_0 \ \ \ \delta_1\right].
  \left[\theta_{0,\infty}(\nu_{1,\infty}) \ \ \ \theta_{1,\infty}(\nu_{1,\infty})\right]^T } \nonumber \\
  && +   \nu_{1,\infty} \theta_{1,\infty}(\nu_{1,\infty}) \geq
              \left[\delta_0 \ \ \ \delta_1\right] \cdot \mathbf{r} + \nu_{1,\infty} r_1 \nonumber \\
\lefteqn{ \nu_{1,\infty} (\theta_{1,\text{min}} - r_1) \geq } \nonumber \\
 && \left[\delta_0 \ \ \ \delta_1\right] \cdot
   \left(\mathbf{r} - \left[\theta_{0,\infty}(\nu_{1,\infty}) \ \ \ \theta_{1,\infty}(\nu_{1,\infty})\right]^T\right)
\end{eqnarray}
where we have used
Equation~(\ref{eqn:stationary-point-index-bias-ode-ue1}) in the left
term of the inequality in the last line of the above sequence of inequalities.

It follows that, for $\mathbf{r} \in \overline{\mathcal{R}}$ with
$r_1 \geq \theta_{1,\text{min}}$, since $\nu_{1,\infty} \geq 0$, we
have shown that (see
Figure~\ref{fig:showing-the-optimality-of-the-ode-limit} for an illustration)
\begin{eqnarray}
  \label{eqn:rg-optimality-condition-is-met}
 \lefteqn{ \left[\frac{d}{d\theta_0}
  U(\theta_{0,\infty}(\nu_{1,\infty})) \  (\frac{d}{d\theta_1}
  U(\theta_{1,\infty}(\nu_{1, \infty})) )\right]. } \nonumber \\
 && \left(\mathbf{r} - \left[\theta_{0,\infty}(\nu_{1,\infty}) \  \theta_{1,\infty}(\nu_{1,\infty})\right]^T\right)
  \leq 0
\end{eqnarray}
Since
$\overline{\mathcal{R}} \cap \{\mathbf{r} \in \mathbb{R}_+: r_1 \geq
\theta_{1,\text{min}}\}$ is a closed convex set, and
$U(\mathbf{r})$ is strictly concave, it follows from
\cite[Chapter 3, Page
103]{books.bazaara-sherali-shetty1993nonlinear-programming} that
$\left[\theta_{0,\infty}(\nu_{1,\infty}) \ \ \
  \theta_{1,\text{min}}\right]$ is the unique utility maximising
throughput vector with a rate guarantee for $\text{UE}_1$.

The above argument suggests that the sequence of EWMA throughputs
will, in large time, become close to
$\left[\theta_{0,\infty}(\nu_{1,\infty}) \ \ \
  \theta_{1,\text{min}}\right]$. This is formalized in the next section.

\subsection{Main result on convergence with rate guarantees}
\label{sec:weak-convergence-with-rate-guarantees}

In this section, under certain conditions, we analyze the algorithm
defined by
Equations~(\ref{eqn:scheduled-rate-vector})-(\ref{eqn:index-bias-update})
using the approach for weak convergence of two-time-scale constant
step-size stochastic approximation provided in \cite[Section
8.6]{books.kushner-yin2003stochastic-approximation-algorithms-with-applications}. We
culminate this section in
Theorem~\ref{thm:convergence-result-formal-statement} which is the
main convergence result for the Lagrange multipliers.

Define the sequence of $\sigma$-fields,
$\mathcal{F}^{(a,b)}_0, \mathcal{F}^{(a,b)}_1, \cdots, \mathcal{F}^{(a,b)}_k, \cdots$, as follows:
\begin{eqnarray*}
  \lefteqn{\mathcal{F}^{(a,b)}_0 = \sigma(\boldsymbol{\theta}^{(a,b)}(0), \boldsymbol{\nu}^{(a,b)}(0)), \cdots,} \\
  && \hspace{-10mm} \mathcal{F}^{(a,b)}_k = \sigma(\boldsymbol{\theta}^{(a,b)}(0), \boldsymbol{\nu}^{(a,b)}(0), s(0), \cdots, \boldsymbol{\theta}^{(a,b)}(k), \boldsymbol{\nu}^{(a,b)}(k))
\end{eqnarray*}

Further, define:
\begin{eqnarray}
  \label{eqn:conditional-mean-function}
\lefteqn{  \boldsymbol{h}(\boldsymbol{\theta}^{(a,b)}(k), \boldsymbol{\nu}^{(a,b)}(k), s(k-1)) :=} \nonumber \\
&&  \mathbb{E} \left(\mathbf{r}^*(\boldsymbol{\theta}^{(a,b)}(k), \boldsymbol{\nu}^{(a,b)}(k), S(k))| \mathcal{F}_k^{(a,b)}\right)
\end{eqnarray}
and observe that $\boldsymbol{h}$ does not depend on $k$ or on $(a,b)$.
% Define
% \begin{eqnarray}
%   \label{eqn:asymptotic-mean-function}
%    \overline{\boldsymbol{h}}^{(a,b)}(\boldsymbol{\theta}, \boldsymbol{\nu}) &:=&
%   \mathbb{E}_\infty \left(r^*(\boldsymbol{\theta}, \boldsymbol{\nu}, S)\right)
% \end{eqnarray}
% where $\mathbb{E}_\infty \left(\cdot\right)$ denotes the expectation
% over the stationary distribution of the channel state Markov process.
The so-called martingale difference ``noise'' in the rate recursion~\eqref{eqn:ewma-rate-update} is defined by:
\begin{eqnarray}
  \label{eqn:martingale-difference-noise}
  \boldsymbol{M}^{(a,b)}(k)& := & \mathbf{r}^*(\boldsymbol{\theta}^{(a,b)}(k), \boldsymbol{\nu}^{(a,b)}(k), s(k)) \nonumber \\
  & & -
     \boldsymbol{h}(\boldsymbol{\theta}^{(a,b)}(k), \boldsymbol{\nu}^{(a,b)}(k), s(k-1)).
\end{eqnarray}
For the index-bias recursion \eqref{eqn:index-bias-update}, we define
\begin{eqnarray}
  \label{eqn:index-bias-mean-function}
  \mathbf{g}(\boldsymbol{\theta}) &:=& \boldsymbol{\theta}_{\text{min}} - \boldsymbol{\theta}
\end{eqnarray}
with the associated martingale difference noise being zero. With these definitions, we rewrite
Equations~(\ref{eqn:ewma-rate-update})-(\ref{eqn:index-bias-update}) as follows:
\begin{eqnarray}
  \label{eqn:ewma-rate-update-rewritten}
\boldsymbol{\theta}^{(a,b)}(k+1) & = & \boldsymbol{\theta}^{(a,b)}(k)  \nonumber \\
  && 
  + a\Big( \boldsymbol{h}(\boldsymbol{\theta}^{(a,b)}(k), \boldsymbol{\nu}^{(a,b)}(k), s(k-1)) \nonumber \\ 
  & & 
  \hspace*{.7cm} - \boldsymbol{\theta}^{(a,b)}(k)
  + \mathbf{M}^{(a,b)}(k) \Big) \\
  \label{eqn:index-bias-update-rewritten}
  \boldsymbol{\nu}^{(a,b)}(k+1) & = & \left[\boldsymbol{\nu}^{(a,b)}(k) +
         b\mathbf{g}(\boldsymbol{\theta}^{(a,b)}(k))\right]_0^{\nu_{\text{max}}}
\end{eqnarray}

The result we will apply is \cite[Theorem
8.6.1]{books.kushner-yin2003stochastic-approximation-algorithms-with-applications}. To apply this result, 
we proceed to verify that the conditions A8.6.0 to A8.6.5 (in
\cite[Section
8.6.1]{books.kushner-yin2003stochastic-approximation-algorithms-with-applications})
hold in our problem.

\textbf{A8.6.0} The projection applied to the index-bias
  $\boldsymbol{\nu}^{(a,b)}(k)$ is on a hyperrectangle, of the
  type described in \cite[A
  4.3.1]{books.kushner-yin2003stochastic-approximation-algorithms-with-applications}).

\textbf{A8.6.1} We notice that $\mathcal{R}_s \subset \mathbb{R}^M_+$
  are compact sets, and
  $\mathbf{r}^*(\boldsymbol{\theta}, \boldsymbol{\nu}, s(k)) \in
  \mathcal{R}_{s(k)} \subset \hat{\mathcal{R}}$ for each $k$. Further,
  by rewriting Equation~(\ref{eqn:ewma-rate-update}) as follows,
  \begin{eqnarray}
    \label{eqn:ewma-recursion-as-convex-combination}
  \lefteqn{  \boldsymbol{\theta}^{(a,b)}(k+1) =} \nonumber \\
  &&\hspace{-10mm}  (1 - a) \boldsymbol{\theta}^{(a,b)}(k) +
    a \mathbf{r}^*(\boldsymbol{\theta}^{(a,b)}(k), \boldsymbol{\nu}^{(a,b)}(k), s(k)),
  \end{eqnarray}
  we observe that
  $\boldsymbol{\theta}^{(a,b)}(k+1)$ will always be in a compact set, and
  uniform integrability of the family of random vectors
  $\{(\mathbf{r}^*(\boldsymbol{\theta}^{(a,b)}(k),
  \boldsymbol{\nu}^{(a,b)}(k), s(k)) -
  \boldsymbol{\theta}^{(a,b)}(k)), (a,b), k\}$, where the random vectors are indexed by $(a,b)$ and $k$, holds. It follows that
  \small
  \begin{eqnarray*}
  \Big\{ \boldsymbol{h}(\boldsymbol{\theta}^{(a,b)}(k),
    \boldsymbol{\nu}^{(a,b)}(k), s(k-1)) - \boldsymbol{\theta}^{(a,b)}(k),
    \boldsymbol{g}(\boldsymbol{\theta}^{(a,b)}(k)), \nonumber \\
    (a,b), k \Big\}
  \end{eqnarray*}
  \normalsize
  is uniformly integrable, and moreover, for each fixed $\boldsymbol{\theta}$ and $\boldsymbol{\nu}$,
  \small
  \begin{eqnarray*}
  \Big\{ \boldsymbol{h}(\boldsymbol{\theta},
    \boldsymbol{\nu}, s(k-1)) - \boldsymbol{\theta},
    \boldsymbol{g}(\boldsymbol{\theta}), (a,b), k \Big\}
  \end{eqnarray*}
  \normalsize
  is also uniformly integrable.

  We further require that the sequence of random variables $S(k), k \geq 0$ is
  tight.  Condition~\ref{cond:compact-channel-state-space} below implies
  this.

  \begin{condition}
    \label{cond:compact-channel-state-space}
  Assume that the channel state process $S(k), k \geq 0,$ belongs
  to a compact set in $\mathbb{R}^M$. \hfill $\Box$
  \end{condition}

  \textbf{A8.6.2} We need the continuity of
  $(\boldsymbol{h}(\boldsymbol{\theta}, \boldsymbol{\nu}, s) -
  \boldsymbol{\theta})$ uniformly over $(a,b)$ and $s$.  Here, the
  $(a,b)$ do not matter; we need the continuity of
  $(h(\boldsymbol{\theta}, \boldsymbol{\nu}, s) -
  \boldsymbol{\theta})$ over $\boldsymbol{\theta}, \boldsymbol{\nu}$
  uniformly in $s$.

  Following \cite{winet.agrawal-subramanian2002optimality-channel-aware-scheduling}, we require:
  \begin{condition}[Strict Convexity of Rate Regions]
    \label{cond:strict-convexity-rate-regions}
    Assume that for every $\mathbf{a} \in \mathbb{R}_+^M$ , and
    $s \in \mathcal{S}$,
    $\arg \max_{\mathbf{r} \in \mathcal{R}_{\cdot|s}}
    \mathbf{a}\cdot\mathbf{r}$ is unique, and, also,
    $\arg \max_{\mathbf{r} \in \overline{\mathcal{R}}}
    \mathbf{a}\cdot\mathbf{r}$ is unique. \hfill $\Box$
  \end{condition}
  The strict convexity is understood to apply to the Pareto optimal boundary of the rate regions.
  
  Further, we require:
  \begin{condition}
    \label{cond:continuity-arg-max-over-rate-regions}
    Assume that for
    $(\mathbf{a},s) \in \mathbb{R}_+^M \times \mathcal{S}$, the
    mappings
    $(\mathbf{a},s) \mapsto \arg \max_{\mathbf{r} \in
      \overline{\mathcal{R}}_{\cdot|s}} \mathbf{a}\cdot\mathbf{r}$ and
    $\mathbf{a} \mapsto \arg \max_{\mathbf{r} \in
      \overline{\mathcal{R}}} \mathbf{a}\cdot\mathbf{r}$ are
    continuous. \hfill $\Box$
  \end{condition}

  % \begin{condition}
  %   \label{cond:smoothness-rate-region-boundary}
  %   Assume that, for each $s \in \mathcal{S}$, the conditional
  %   expected rate region $\overline{\mathcal{R}}_{\cdot|s}$ is such
  %   that each point on its Pareto boundary has a unique supporting
  %   hyperplane. Further, assume that this also holds for the expected
  %   rate region $\overline{\mathcal{R}}$. \hfill $\Box$
  % \end{condition}

  If the $\mathcal{R}_s$ is polyhedral for each $s \in \mathcal{S}$,
  Condition 3 may hold if $\mathcal{S}$ is a continuum\footnote{In the
    sequel, we will continue to use the notation
    $\sum_{s\in \mathcal{S}} (\cdot)$ and $P(\cdot|s)$ with the
    understanding that these should be intrepreted as integrals and
    transition kernel densities, respectively, when $\mathcal{S}$ is a
    continuum.}, for example when we are dealing with Rayleigh fading
  with truncation to a compact set (in view of Condition
  \ref{cond:compact-channel-state-space}).

  Analogous
  to~\eqref{eqn:equivalence-of-average-state-optimum-and-global-optimum},
  using Condition~\ref{cond:strict-convexity-rate-regions}, it is
  easily seen that (see
  \cite{winet.agrawal-subramanian2002optimality-channel-aware-scheduling}):
  \begin{eqnarray}
  \label{eqn:equivalence-of-average-state-optimum-and-global-optimum-2}
  \arg \max_{\mathbf{r} \in \overline{\mathcal{R}}_{\cdot | s}} \mathbf{a}^T\cdot\mathbf{r} &=&
   \sum_{s \in \mathcal{S}} P(s^\prime | s)  \arg \max_{\mathbf{r} \in \mathcal{R}_{s^\prime}} \mathbf{a}^T\cdot\mathbf{r}
  \end{eqnarray}
  Since $(\nabla U(\boldsymbol{\theta}) + \boldsymbol{\nu})^T$ is
  continuous in $(\boldsymbol{\theta}, \boldsymbol{\nu})$, Condition \ref{cond:continuity-arg-max-over-rate-regions} and the compactness of $\mathcal{S}$ implies that
  $\boldsymbol{h}(\boldsymbol{\theta}, \boldsymbol{\nu}, s) -
  \boldsymbol{\theta}$ is continuous over
  $(\boldsymbol{\theta}, \boldsymbol{\nu})$ uniformly in $s$.
  %\textbf{ Continuity in
  %$\boldsymbol{\theta}, \boldsymbol{\nu}$ uniformly over the channel
  %  states $s$ will require a proof or additional assumptions. The
  %  joint continuity of
  %  $\mathbf{r}^*(\boldsymbol{\theta}, \boldsymbol{\nu}, s)$ in
  %  $\boldsymbol{\theta}, \boldsymbol{\nu}, s$, over compact
  %  $(\overline{\mathcal{R}}, [0, \nu_{\text{max}}]^M, \mathcal{S})$ will
  %  suffice.}

  Trivially, $\mathbf{g}(\boldsymbol{\theta})$ is continuous
  in $\boldsymbol{\theta}$.

  \textbf{A8.6.3} Fix $(\boldsymbol{\theta}, \boldsymbol{\nu})$. Consider the following (notice that the step sizes $(a,b)$ do not play any role) for a suitable continuous function $\overline{\mathbf{h}}(\boldsymbol{\theta}, \boldsymbol{\nu})$:
  \small
  \begin{eqnarray*}
    \lefteqn{ \lim_{\stackrel[\frac{b}{a} \to 0]{n, m \to \infty}{\tiny{a \to 0}}}
    \frac{1}{m}  \hspace{-1.5ex} \sum_{i=n}^{n + m -1} \hspace{-2ex}\left( \mathbb{E} \left(\mathbf{h}(\boldsymbol{\theta}, \boldsymbol{\nu}, S(i-1)) - \boldsymbol{\theta}| \mathcal{F}_n^{(a,b)}\right) - \overline{\mathbf{h}}(\boldsymbol{\theta}, \boldsymbol{\nu})\right)} \\
    && =  \lim_{(n, m \to \infty)}
       \frac{1}{m}  \hspace{-1.5ex} \sum_{i=n}^{n + m -1}  \hspace{-2ex} \left( \mathbb{E} \left(\mathbf{r}^*(\boldsymbol{\theta}, \boldsymbol{\nu}, S(i))  - \boldsymbol{\theta}|
       \mathcal{F}_n^{(a,b)}\right) - \overline{\mathbf{h}}(\boldsymbol{\theta}, \boldsymbol{\nu})\right) \\
    && = \lim_{(m \to \infty)}
       \frac{1}{m} \sum_{i=1}^{m} \left(\mathbb{E} \left(\mathbf{r}^*(\boldsymbol{\theta}, \boldsymbol{\nu}, S(i))  - \boldsymbol{\theta}| s(0)\right) - \overline{\mathbf{h}}(\boldsymbol{\theta}, \boldsymbol{\nu})\right) \\
    && = \mathbb{E}_\infty \mathbf{r}^*(\boldsymbol{\theta}, \boldsymbol{\nu}, S)  - \boldsymbol{\theta} - \overline{\mathbf{h}}(\boldsymbol{\theta}, \boldsymbol{\nu})
  \end{eqnarray*}
  \normalsize in probability, where we have used ergodicity of the
  Markov chain of channel states and the uniform integrability of
  $\left\{\left(\mathbf{r}^*(\boldsymbol{\theta}, \boldsymbol{\nu},
      S(i)) | s(0)\right), i \geq 1 \right\}$ in the last equality and
  $\mathbb{E}_{\infty}(\cdot)$ refers to the expectation with respect
  to the stationary distribution of the Markov chain. It follows that,
  for the last term in the above sequence of equalities to be equal to
  $0$,
  \begin{eqnarray}
    \label{eqn:ewma-ode-with-fixed-index-bias-driving-term}
 \overline{\mathbf{h}}(\boldsymbol{\theta}, \boldsymbol{\nu}) &=& \mathbb{E}_\infty \mathbf{r}^*(\boldsymbol{\theta}, \boldsymbol{\nu}, S)  - \boldsymbol{\theta}\nonumber \\
&=& \arg \max_{\boldsymbol{r} \in \overline{\mathcal{R}}} (\nabla U(\boldsymbol{\theta}) +  \boldsymbol{\nu}^T)\cdot\mathbf{r}  - \boldsymbol{\theta}
  \end{eqnarray}
  is the desired function. The second equality requires
  Condition~\ref{cond:strict-convexity-rate-regions}, and the
  continuity of
  $\overline{\mathbf{h}}(\boldsymbol{\theta}, \boldsymbol{\nu})$
  with respect to $(\boldsymbol{\theta}, \boldsymbol{\nu})$ follows from
  Condition~\ref{cond:continuity-arg-max-over-rate-regions}.

%  \begin{condition}
%    \label{cond:strictly-convex-rate-regions}
%    For every $s \in \mathcal{S}$, the rate region $\mathcal{R}_s$ is
%    strictly convex, and so is $\overline{\mathcal{R}}$. \hfill $\Box$
%  \end{condition}
%    % \textbf{Check: The second assertion is implied by the first one,
%    %   if the convex combination of open balls is open?}
%
%

\textbf{A8.6.4} We now need to show that, for fixed $\boldsymbol{\nu}$,
  the o.d.e.
  \begin{eqnarray}
    \label{eqn:ewma-ode-with-fixed-index-bias}
    \dot{\boldsymbol{\theta}}(t) &=&
       \overline{\mathbf{h}}(\boldsymbol{\theta}(t), \boldsymbol{\nu})
  \end{eqnarray}
  has a unique globally asymptotically stable point that is continuous in
  $\boldsymbol{\nu}$, characterized by
    \begin{eqnarray}
    \label{eqn:ewma-ode-fixed-index-bias-stable-points}
      \boldsymbol{\theta}_\infty(\boldsymbol{\nu}) &=&
     \arg \max_{\mathbf{r} \in \overline{\mathcal{R}}}(\nabla U(\boldsymbol{\theta}_\infty(\boldsymbol{\nu})) +
    \boldsymbol{\nu})^T\cdot\mathbf{r}
  \end{eqnarray}

  We first argue that the rest point of the dynamics~\eqref{eqn:ewma-ode-with-fixed-index-bias} is unique for 
  each $\boldsymbol{\nu} \in \mathbb{R}^M_+$, and that the rest point is
  $\boldsymbol{\theta}_\infty(\boldsymbol{\nu})$. Consider the
  optimization problem
  $\max_{\mathbf{r} \in \overline{\mathcal{R}}} (U(\mathbf{r}) +
  \boldsymbol{\nu}^T\cdot \mathbf{r})$. This is the maximization of a
  strictly concave function over a convex set; hence, the solution,
  say $\mathbf{r}^*(\boldsymbol{\nu})$, is unique. Further, this
  optimum solution must satisfy
  $ \left(\nabla U(\mathbf{r}^*(\boldsymbol{\nu})) +
    \boldsymbol{\nu}\right)^T\cdot(\mathbf{r} -
  \mathbf{r}^*(\boldsymbol{\nu})) \leq 0$ for all
  $\mathbf{r} \in \overline{\mathcal{R}}$; i.e.,
  $ \mathbf{r}^*(\boldsymbol{\nu}) = \arg \max_{\mathbf{r} \in
  \overline{\mathcal{R}}}\left(\nabla U(\mathbf{r}^*(\boldsymbol{\nu})) +
  \boldsymbol{\nu}\right)^T\cdot\mathbf{r}$. Observe from~\eqref{eqn:ewma-ode-fixed-index-bias-stable-points} that this $\mathbf{r}^*(\boldsymbol{\nu})  = \boldsymbol{\theta}_\infty(\boldsymbol{\nu})$, so the right-hand-side of~\eqref{eqn:ewma-ode-with-fixed-index-bias-driving-term} is zero at $\boldsymbol{\theta} =  \boldsymbol{\theta}_\infty(\boldsymbol{\nu})$, and hence $\boldsymbol{\theta}_\infty(\boldsymbol{\nu})$ is the unique rest point for the o.d.e.~\eqref{eqn:ewma-ode-with-fixed-index-bias}.

  In the Appendix, we show that
  $\boldsymbol{\theta}_\infty(\boldsymbol{\nu})$ is globally asymptotically stable for the o.d.e.~(\ref{eqn:ewma-ode-with-fixed-index-bias}).

%  Returning to
%  Equation~\ref{eqn:ewma-ode-fixed-index-bias-stable-points}, we
%  conclude that, for each $\boldsymbol{\nu} \in \mathbb{R}^M_+$,
%  $\boldsymbol{\theta}_\infty(\boldsymbol{\nu})$ is the unique
%  globally asymptotically stable point of the o.d.e.\ displayed in
%  Equation~(\ref{eqn:ewma-ode-with-fixed-index-bias}).

  Continuity of the unique stable point
  $\boldsymbol{\theta}_\infty(\boldsymbol{\nu})$ with
  $\boldsymbol{\nu}$: Continuing the previous argument, we have
  $ \mathbf{r}^*(\boldsymbol{\nu}) = \arg \max_{\mathbf{r} \in
    \overline{\mathcal{R}}}\left(\nabla
    U(\mathbf{r}^*(\boldsymbol{\nu})) +
    \boldsymbol{\nu}\right)^T\cdot\mathbf{r}$. Consider the sequence,
  indexed by $k \geq 1$,
  $\boldsymbol{\nu}_k \to_{k \to \infty} \boldsymbol{\nu}_0$ in
  $\mathbb{R}^M_+$. For every $k \geq 1$,
  $\mathbf{r}^*(\boldsymbol{\nu_k}) \in \overline{\mathcal{R}}$, a
  compact set. Thus, consider a subsequence
  $\mathbf{r}^*(\boldsymbol{\nu}_{k_j}) \to_{j \to \infty}
  \mathbf{r}^\prime$ and
  $\boldsymbol{\nu}_{k_j} \to_{j \to \infty}
  \boldsymbol{\nu}_0$. Since $U(\boldsymbol{\theta})$ is continuously
  differentiable,
  $\nabla U(\mathbf{r}^*(\boldsymbol{\nu}_{k_j})) +
  \boldsymbol{\nu}_{k_j} \to_{j \to \infty} \nabla
  U(\mathbf{r}^\prime) + \boldsymbol{\nu}_0$. Hence, by
  Condition~\ref{cond:strict-convexity-rate-regions},
  \begin{eqnarray*}
    \lefteqn{ \lim_j \arg \max_{\mathbf{r} \in \overline{\mathcal{R}}}\left(\nabla
    U(\mathbf{r}^*(\boldsymbol{\nu}_{k_j})) +
    \boldsymbol{\nu}_{k_j}\right)^T\cdot\mathbf{r}} \\
    & =&
    \arg \max_{\mathbf{r} \in \overline{\mathcal{R}}}\left(\nabla
    U(\mathbf{r}^\prime) +
    \boldsymbol{\nu}_0\right)^T\cdot\mathbf{r};
  \end{eqnarray*}
  i.e.,
  $\mathbf{r}^*(\boldsymbol{\nu}_{k_j}) \to_{j \to \infty} \arg
  \max_{\mathbf{r} \in \overline{\mathcal{R}}}\left(\nabla U(\mathbf{r}^\prime) +
    \boldsymbol{\nu}_0\right)^T\cdot\mathbf{r}$, or
  $\mathbf{r}^\prime = \arg \max_{\mathbf{r} \in
    \overline{\mathcal{R}}}\left(\nabla U(\mathbf{r}^\prime) +
    \boldsymbol{\nu}_0\right)^T\cdot\mathbf{r}$. By the uniqueness,
  proved above,
  $\mathbf{r}^\prime = \mathbf{r}^*(\boldsymbol{\nu}_0)$. This holds
  for every convergent subsequence of
  $\{\mathbf{r}^\prime(\boldsymbol{\nu}_k), k \geq 1\}$. Hence,
  $\textbf{r}^*(\boldsymbol{\nu}_k) \to_{k \to \infty}
  \textbf{r}^*(\boldsymbol{\nu}_0)$, and we conclude that $\boldsymbol{\theta}_\infty(\boldsymbol{\nu})$ is
  continuous in $\boldsymbol{\nu}$.

\textbf{A8.6.5} We first need to show uniform integrability of the sequence
  $\{\mathbf{g}(\boldsymbol{\theta}_\infty(\boldsymbol{\nu}^{(a,b)}(k))),
  (a,b), k\}$, and, for each $\boldsymbol{\nu}$, of the sequence
  $\{\mathbf{g}(\boldsymbol{\theta}_\infty(\boldsymbol{\nu})),
  (a,b)\}$. The first follows because the sequence belong to a compact set. The second is trivial because $\mathbf{g}$ is independent of $(a,b)$.

  Further, we must show that there is a continuous function
  $\boldsymbol{\nu} \mapsto \overline{\mathbf{g}}(\boldsymbol{\theta}_\infty(\boldsymbol{\nu}),
  \boldsymbol{\nu})$ such that for each $\boldsymbol{\nu}$, in
  probability, the following holds:
  \small
  \begin{eqnarray*}
 \lefteqn{    \lim_{\stackrel[\frac{b}{a} \to 0]{n, m \to \infty}{\tiny{a \to 0}}}
    \frac{1}{m} \sum_{i=n}^{n + m -1} \hspace{-2ex} \left( \mathbb{E} \left(\mathbf{g}(\boldsymbol{\theta}_\infty(\boldsymbol{\nu}), \boldsymbol{\nu})| \mathcal{F}_n^{(a,b)}\right) -
    \overline{\mathbf{g}}(\boldsymbol{\theta}_\infty(\boldsymbol{\nu}), \boldsymbol{\nu})\right) } \\
    && \hspace{70mm}= 0
  \end{eqnarray*}
  \normalsize
  i.e.,
  \small
 \begin{eqnarray*}
  \lim_{\stackrel[\frac{b}{a} \to 0]{n, m \to \infty}{\tiny{a \to 0}}}
    \frac{1}{m} \sum_{i=n}^{n + m -1} (\boldsymbol{\theta}_{\text{min}} - \boldsymbol{\theta}_\infty(\boldsymbol{\nu})) -
    \overline{\mathbf{g}}(\boldsymbol{\theta}_\infty(\boldsymbol{\nu}), \boldsymbol{\nu}) = 0
  \end{eqnarray*}
  \normalsize
It follows that the desired function is
  \begin{eqnarray}
    \label{eqn:index-bias-ode-driving-function}
    \overline{\mathbf{g}}(\boldsymbol{\theta}_\infty(\boldsymbol{\nu}), \boldsymbol{\nu}) &=& \boldsymbol{\theta}_{\text{min}} - \boldsymbol{\theta}_\infty(\boldsymbol{\nu})
  \end{eqnarray}
  which is continuous in $\boldsymbol{\nu}$ by A8.6.4.

The following condition on the
optimization problem in Equation~(\ref{eqn:optimization-problem})
ensures that we can restrict $\boldsymbol{\nu}$ to a compact set:
\begin{condition}
  \label{cond:nu-max-existence}
  There is a $\nu_{\text{max}} > 0$ such that the optimum Lagrange
  multipliers (dual variables) of the optimization problem
  (\ref{eqn:optimization-problem}) lie in the set
  $[0, \nu_{\text{max}})^M$. \hfill $\Box$
\end{condition}
In the Appendix, we illustrate how this condition can be checked, by
providing an example of a two UE problem.

With all the above conditions having been met, along with the o.d.e.\
in Equation~(\ref{eqn:ewma-ode-with-fixed-index-bias}), \cite[Theorem
8.6.1]{books.kushner-yin2003stochastic-approximation-algorithms-with-applications}
considers the following o.d.e.\ for the index-bias.
\begin{eqnarray}
  \label{eqn:index-bias-ode-in-the-limit}
  \dot{\boldsymbol{\nu}}(t) &=& \boldsymbol{\theta}_{\text{min}} - \boldsymbol{\theta}_\infty(\boldsymbol{\nu}(t)) + \boldsymbol{\xi}(t), \quad t \geq 0
\end{eqnarray}
where $\boldsymbol{\xi}(t) \in -C(\boldsymbol{\nu}(t))$ is the term
that compensates for the dynamics so that $\boldsymbol{\nu}(t)$ stays
in its constraint set; here $C(\boldsymbol{\nu})$ is the cone of
outward normals to the constraint set at $\boldsymbol{\nu}$.

Let $L$ denote the set of limit points of the
dynamics~\eqref{eqn:index-bias-ode-in-the-limit} in $[0,\nu_{\max}]^M$
over all initial conditions. For a $\delta > 0$, let $L^{(\delta)}$
denote the $\delta$-neighborhood of $L$. Define the continuous time
interpolation $\hat{\boldsymbol{\nu}}^{(a,b)}(t)$ for $t \geq 0$ as:
\[
  \hat{\boldsymbol{\nu}}^{(a,b)}(t) = \boldsymbol{\nu}^{(a,b)}(k), \quad t \in [bk, b(k+1)).
\]
The foregoing has established the following theorem.

\begin{theorem}
  \label{thm:convergence-result-formal-statement}
  Let $q_a$ be any sequence of integers such that $a q_a \rightarrow \infty$ as $a \rightarrow 0$. For the model considered in this paper, under
  Conditions~\ref{cond:compact-channel-state-space}, \ref{cond:strict-convexity-rate-regions},
  \ref{cond:continuity-arg-max-over-rate-regions}, and
  \ref{cond:nu-max-existence}, there exist $T_a \rightarrow \infty$ as $a \rightarrow 0$ such that for every $\delta > 0$, we have
  \begin{equation}
    \lim_{\stackrel{b/a \rightarrow 0}{a \rightarrow 0}} P \left\{ \hat{\boldsymbol{\nu}}^{(a,b)}(aq_a + t) \notin L^{(\delta)}, \text{ for some } t \leq T_a \right\} = 0.
  \end{equation}
\end{theorem}

\begin{IEEEproof}
 All assumptions A8.6.0 to A.8.6.5, in order to apply \cite[Theorem 8.6.1]{books.kushner-yin2003stochastic-approximation-algorithms-with-applications}, have been verified above, and the theorem follows.
\end{IEEEproof}

\subsection{Stable rest points and the optimum solution}
\label{sec:stable-points-and-the-optimum-solution}

We now use properties of the joint stable points of the
o.d.e. in Equation~(\ref{eqn:ewma-ode-with-fixed-index-bias}) and the
o.d.e. in Equation~(\ref{eqn:index-bias-ode-in-the-limit}), i.e.,
$(\boldsymbol{\theta}_\infty(\boldsymbol{\nu}_\infty),
\boldsymbol{\nu}_\infty)$, to show that these correspond to the
optimal solution of the optimization problem in
Equation~(\ref{eqn:optimization-problem}).

Consider a stable rest point $\boldsymbol{\nu}_\infty$ of the o.d.e.\
in Equation~(\ref{eqn:index-bias-ode-in-the-limit}). Such a rest point
is in $L$.  Using
Equation~(\ref{eqn:ewma-ode-fixed-index-bias-stable-points}), we can
write, for every $\mathbf{r} \in \overline{\mathcal{R}}$:
\begin{eqnarray*}
  (\nabla U(\boldsymbol{\theta}_\infty(\boldsymbol{\nu}_\infty)) +
  \boldsymbol{\nu}_\infty)^T \cdot \left(\boldsymbol{\theta}_\infty(\boldsymbol{\nu}_\infty) -   \mathbf{r}\right) &\geq& 0
\end{eqnarray*}
Transposing, we obtain, for every $\mathbf{r} \in \overline{\mathcal{R}}$:
\begin{eqnarray}
  \label{eqn:optimality-check-using-complementary-slackness}
   \lefteqn{\boldsymbol{\nu}_\infty^T \cdot (\boldsymbol{\theta}_\infty(\boldsymbol{\nu}_\infty) - \mathbf{r})  \geq} \nonumber \\
  && (\nabla U(\boldsymbol{\theta}_\infty(\boldsymbol{\nu}_\infty)))^T.(\mathbf{r} - \boldsymbol{\theta}_\infty(\boldsymbol{\nu}_\infty))
\end{eqnarray}
Since $\boldsymbol{\nu}_\infty$ is stable rest point for the o.d.e.\
in Equation~(\ref{eqn:index-bias-ode-in-the-limit}), and
$\boldsymbol{\xi}(t) \in -C(\boldsymbol{\nu}(t))$, it follows that,
for $1 \leq i \leq M$ (the number of UEs):
\begin{eqnarray}
  \label{eqn:index-bias-complementary-slackness}
  (\boldsymbol{\nu}_\infty)_i > 0 \ &\Rightarrow& \
                               (\boldsymbol{\xi}_\infty)_i = 0 \ \ \ \text{and} \ \ \
                               \theta_{i,\text{min}}= (\boldsymbol{\theta}_\infty(\boldsymbol{\nu}_\infty))_i \nonumber \\
  (\boldsymbol{\nu}_\infty)_i = 0 \ &\Rightarrow& \
                               (\boldsymbol{\xi}_\infty)_i \ne 0 \ \ \ \text{and} \ \ \
                               \theta_{i,\text{min}}< (\boldsymbol{\theta}_\infty(\boldsymbol{\nu}_\infty))_i \nonumber \\
\end{eqnarray}
Note that the case
$(\boldsymbol{\nu}_\infty)_i = 0, \theta_{i,\text{min}}>
(\boldsymbol{\theta}_\infty(\boldsymbol{\nu}_\infty))_i$ cannot be a
stable rest point of the o.d.e.\
Equation~(\ref{eqn:index-bias-ode-in-the-limit}).

Returning to
Equation~\eqref{eqn:optimality-check-using-complementary-slackness}, and
utilising Equations~\eqref{eqn:index-bias-complementary-slackness}, we
conclude that for
$\mathbf{r} \in \overline{\mathcal{R}} \cap \{\mathbf{r} \in \mathbb{R}^M_+:
\mathbf{r} \geq \boldsymbol{\theta}_{\text{min}}\}$,
\begin{eqnarray*}
  0 &\geq&
   (\nabla U(\boldsymbol{\theta}_\infty(\boldsymbol{\nu}_\infty)))^T \cdot (\mathbf{r} - \boldsymbol{\theta}_\infty(\boldsymbol{\nu}_\infty))
\end{eqnarray*}
From \cite[Chapter 3, Page
103]{books.bazaara-sherali-shetty1993nonlinear-programming}, we see
that
$(\boldsymbol{\theta}_\infty(\boldsymbol{\nu}_\infty),
\boldsymbol{\nu}_\infty )$ are optimal for the optimization problem in
Equation~(\ref{eqn:optimization-problem}). Since the optimization is
one of a strictly concave function over a convex set, the solution is
unique, and $\boldsymbol{\nu}_{\infty}$ is the correct set of Lagrange multipliers.

Thus
$(\boldsymbol{\theta}_\infty(\boldsymbol{\nu}_\infty),
\boldsymbol{\nu}_\infty)$ is a joint unique stable rest point for the
joint o.d.e. system in
Equation~(\ref{eqn:ewma-ode-with-fixed-index-bias}), with
$\boldsymbol{\nu}$ replaced by $\boldsymbol{\nu}(t)$, and
Equation~(\ref{eqn:index-bias-ode-in-the-limit}).

We have already seen, in the verification of A8.6.4, that, for fixed
index-bias $\boldsymbol{\nu}$, the EWMA rate recursion yields the
o.d.e.\ in Equation~(\ref{eqn:ewma-ode-with-fixed-index-bias}) which
has the unique globally asymptotically stable point
$\boldsymbol{\theta}_\infty(\boldsymbol{\nu})$, characterized in
Equation~(\ref{eqn:ewma-ode-fixed-index-bias-stable-points}), and
continuous in $\boldsymbol{\nu}$. Further, the index-bias recursion
yields the o.d.e.\ in
Equation~(\ref{eqn:index-bias-ode-in-the-limit}). Theorem
\ref{thm:convergence-result-formal-statement} establishes that if, in
addition, $L = \{\boldsymbol{\nu}_{\infty}\}$, then, for large $k$,
$(\boldsymbol{\theta}^{(a,b)}(k), \boldsymbol{\nu}^{(a,b)}(k))$ are
close to
$(\boldsymbol{\theta}_\infty(\boldsymbol{\nu}_\infty),
\boldsymbol{\nu}_\infty)$, which solve the optimization problem in
Equation~(\ref{eqn:optimization-problem}). In other words, the
throughput iterates converge to the optimum solution. A note of
caution however is that $L$ could contain limit cycles for the
dynamics in~\eqref{eqn:index-bias-ode-in-the-limit}. In the simulation
experiments, we see that we do not have such issues on the examples
considered.

%\subsubsection{Convergence of the iterates}
%\label{sec:weak-convergence}
%The preceding sections have established the following which is the main result of this paper.
%\begin{theorem}
%  For the model considered in this paper, under
%  Conditions~\ref{cond:compact-channel-state-space},
%  \ref{cond:continuity-arg-max-over-rate-regions}, and
%  \ref{cond:nu-max-existence} , for every $\varepsilon > 0$, there is
%  a $T$ such that for all $k \geq T$, we have
%\begin{equation}
%  \limsup_{\stackrel{b/a \rightarrow 0}{a \rightarrow 0}} P \left\{ \sup_{k \geq T} \| \boldsymbol{\theta}^{(a,b)}(k)- \boldsymbol{\theta}_{\infty}(\boldsymbol{\nu}_{\infty}) \| > \varepsilon \right\} = 0.
%\end{equation}
%\end{theorem}

\section{Simulation Experiments}
\label{sec:simulation-experiments}

We will carry out simulation experiments for the utility function
$U_i(r_i) = \ln(1 + r_i)$, which satisfies all the mathematical
properties in our earlier development. With a slight abuse of
terminology, we will call the resulting rate allocation
\emph{proportionally fair}, even though the addition of a 1 to the
rate $r_i$ is a deviation from the commonly accepted $\ln{r_i}$. For
modern wireless networks, the rates available to UEs are very high
(several 10s of Mbps); the addition of the 1 will not result in a
significant difference in the final rate allocation, and
$\ln(1 + r_i)$ reduces mathematical complexities. The rate allocation
algorithm with rate guarantees, presented in
Section~\ref{sec:gradient-scheduling-rate-guarantees} will be referred
to as PF-RG-LM (Proportional Fair scheduling with Rate Guarantees,
using the Lagrange Multiplier approach).

\subsection{The PF-RG-TC algorithm}
\label{sec:pf-rg-tc-algorithm}

In
\cite{winet.mandelli-andrews-borst-klein2019satisfying-network-slicing-constraints-via-5G-mac-scheduling,
  winet.andrews-mandelli-borst2018networks-slicing-using-token-counters},
Mandelli et al. have also provided an index-bias based algorithm. They
maintain a token counter (denoted, here, by $\tau(k)$) which is used
to bias the rate allocation index. We call their algorithm PF-RG-TC
(for Token Counter). Following the notation in
Equations~(\ref{eqn:scheduled-rate-vector}),
(\ref{eqn:ewma-rate-update}), and (\ref{eqn:index-bias-update}), we
write the PF-RG-TC as follows.

There is a single small step-size $a > 0$. At the beginning of
slot~$k$, given the joint channel state $s(k) \in \mathcal{S}$, the
EWMA throughputs $\boldsymbol{\theta}^{(a)}(k)$, and the token
counters $\tau_i^{(a)}(k) \geq 0, 1 \leq i \leq M$, determine
$\mathbf{r}^*(\boldsymbol{\theta}^{(a)}(k),
\boldsymbol{\tau}^{(a)}(k), s(k))$ as follows (see \cite[Equations (4)
and
(5)]{winet.mandelli-andrews-borst-klein2019satisfying-network-slicing-constraints-via-5G-mac-scheduling}):

\begin{eqnarray}
  \label{eqn:pf-rg-tc-scheduled-rate-vector}
 \lefteqn{ \mathbf{r}^*(\boldsymbol{\theta}^{(a)}(k), \boldsymbol{\tau}^{(a)}(k), s(k)) =} \nonumber \\
 && \arg \max_{\mathbf{r} \in \mathcal{R}_{s(k)}} (\nabla U(\boldsymbol{\theta}^{(a)}(k)) +
  a \boldsymbol{\tau}^{(a)}(k))^T\cdot\mathbf{r}
\end{eqnarray}

The EWMA throughputs are updated as follows.
\begin{eqnarray}
  \label{eqn:pf-rg-tc-ewma-rate-update}
 \lefteqn{ \boldsymbol{\theta}^{(a)}(k+1) =} \nonumber \\
  && \hspace{-5mm}
   \boldsymbol{\theta}^{(a)}(k) + a(\mathbf{r}^*(\boldsymbol{\theta}^{(a)}(k), \boldsymbol{\tau}^{(a)}(k), s(k)) -
  \boldsymbol{\theta}^{(a)}(k))
\end{eqnarray}

For all $i$, the index biases are updated as follows.
\begin{eqnarray}
  \label{eqn:pf-rg-tc-index-bias-update}
  \lefteqn{ \boldsymbol{\tau}^{(a)}(k+1) =} \nonumber \\
  && \left[\boldsymbol{\tau}^{(a)}(k) + (\boldsymbol{\theta}^{(a)}_{\text{min}} -
  \mathbf{r}^*(\boldsymbol{\theta}^{(a)}(k), \boldsymbol{\tau}^{(a)}(k), s(k)) \right]_0^{\tau_{\text{max}}}
\end{eqnarray}

In Section~\ref{sec:two-ues-comparison-pf-rg-lm}, for two UEs we
provide a performance comparison of PF-RG-LM and PF-RG-TC.  In
Section~\ref{sec:pf-rg-lm-n-UEs} we provide a simulation study of
PF-RG-LM for more than two UEs.

\subsection{Wireless communication model}
\label{sec:wireless-communication-model}

All our simulations are for the following generic wireless
communication setting. The channel is in the 2.4~GHz band, and the
system bandwidth is $W = 40$~MHz. The noise floor is taken as
$\text{Noise} = -97$~dBm; there are no interferers. For the BS
transmit power $P$~dBm, the mean received power at distance $d$~m is
given by $P - 42 - (30 \times \log_{10}d)$, where $42$ is the
attenuation at $1$~m, and the pathloss exponent is $3$. Rayleigh
fading is assumed, which results in exponentially distributed
attuation in dB, with parameter 1. The slot duration is $0.001$~sec.

In slot $k$, the received signal strengths
$\text{RSS}_i(k), i \in \{0,1\},$ are sampled as above. Then, the rates
available to the two UEs are obtained independently (in time and
across the UEs) from the Shannon channel capacity formula as follows:
\begin{eqnarray*}
  r_{i,s(k)} = W \log_2\left(1 + 10^{(\frac{(\text{RSS}_i - \text{Noise})}{10})}\right)
\end{eqnarray*}

\subsection{Two UEs: Rate regions}
\label{sec:two-ues-rate-regions}

\begin{figure}[!h]
\begin{center}
\includegraphics[scale=0.21]{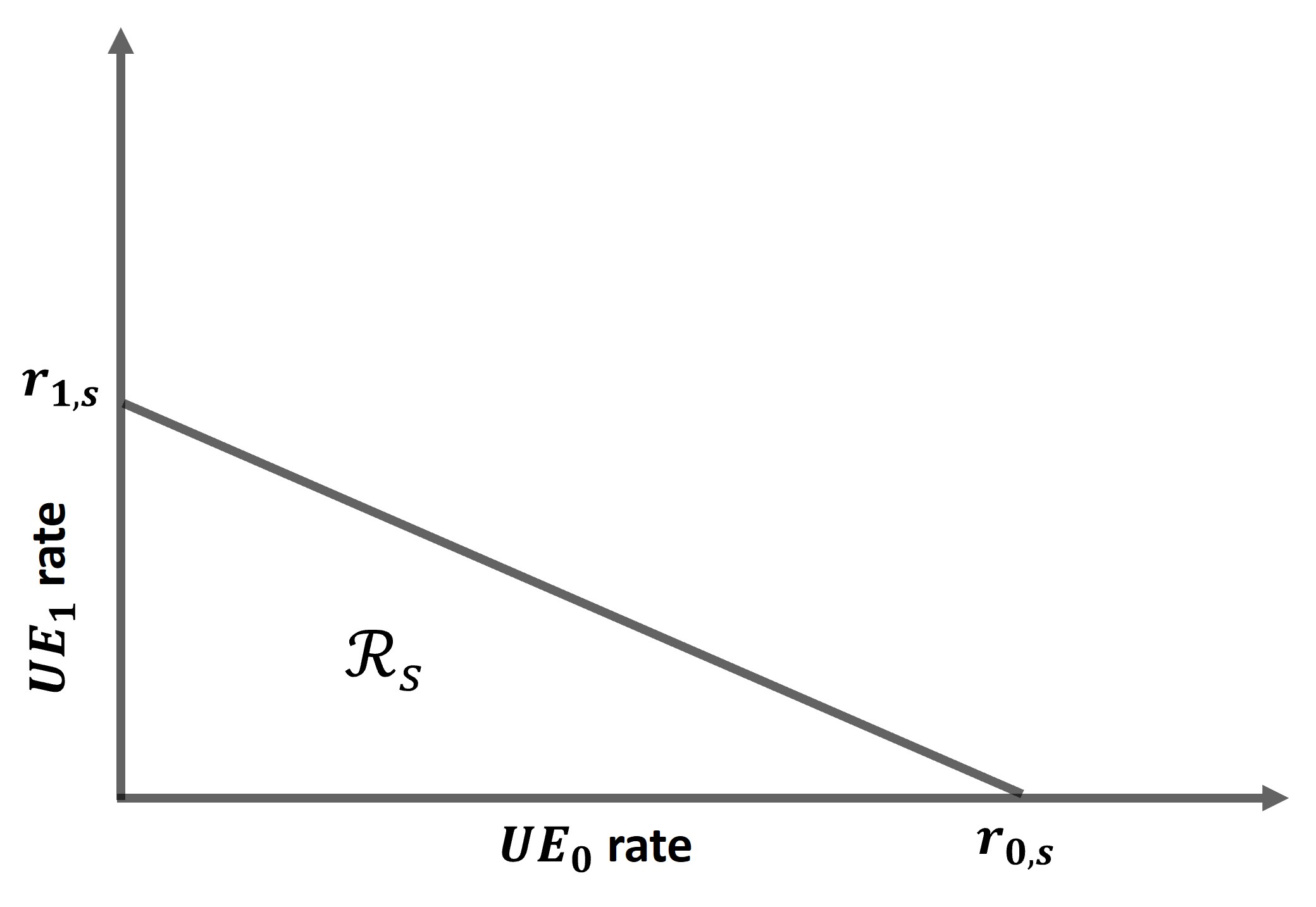}
\hspace{1mm}
\includegraphics[scale=0.06]{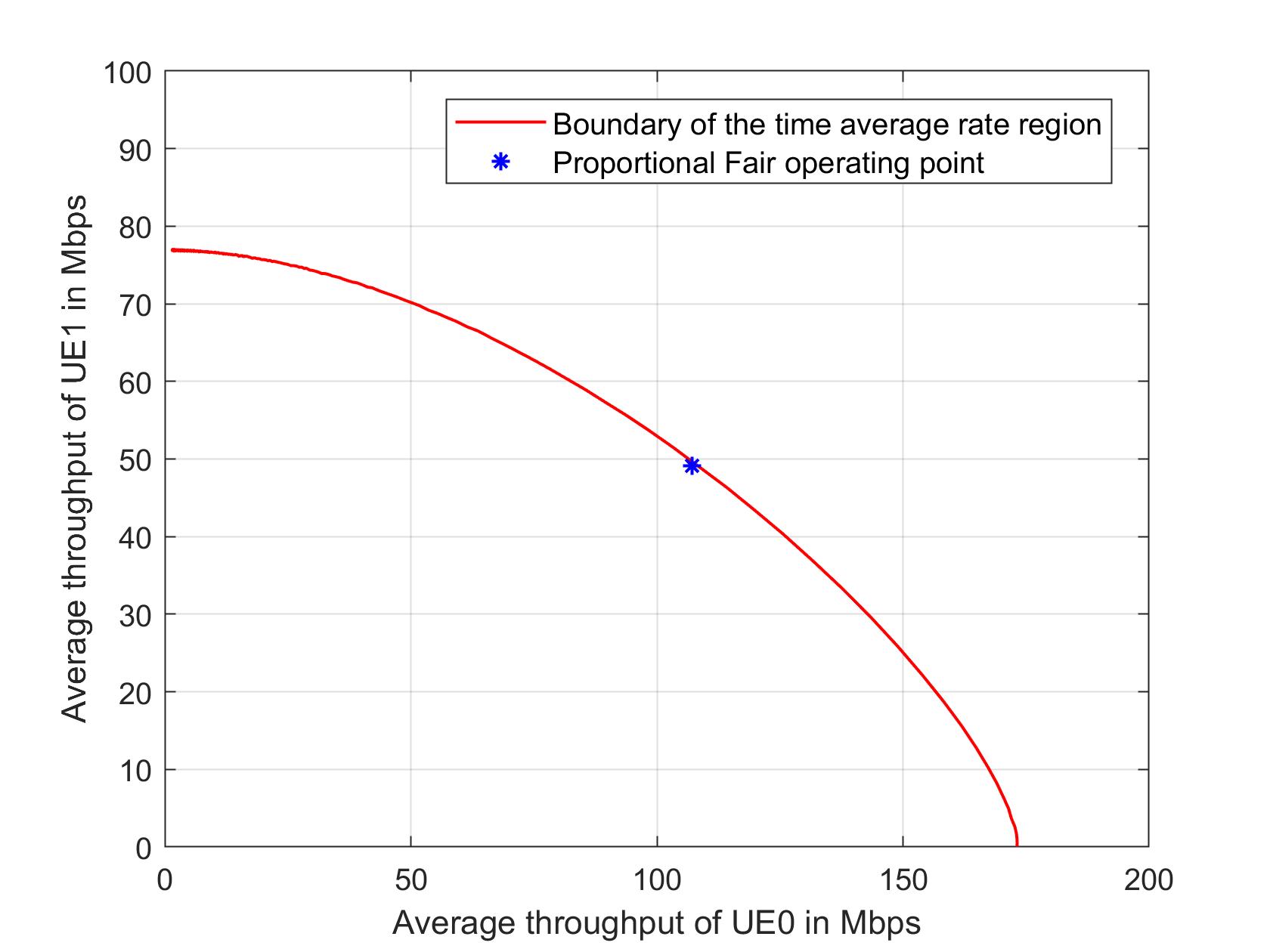}
\end{center}
\caption{Left plot: A slot rate region $\mathcal{R}_s$.  Right plot:
  The rate region $\overline{\mathcal{R}}$.}
\label{fig:two-ues-slot-and-time-average-rate-regions}
\end{figure}

There are two UEs, $\text{UE}_0$ and $\text{UE}_1$, at the distances
of $100$~m and $200$~m from the gNB.  With the wireless channel model
in Section~\ref{sec:wireless-communication-model}, and $P = 20$~dBm,
the rate region in a slot, $\mathcal{R}_s, s \in \mathcal{S}$, has the
triangular shape depicted in the left panel
Figure~\ref{fig:two-ues-slot-and-time-average-rate-regions}. For such
a rate region for two UEs, when the calculation in
Equation~(\ref{eqn:scheduled-rate-vector}) is done, it is of the form
$\arg \max_{(r_0, r_1) \in \mathcal{R}_s} (a_0 r_0 + a_1 r_1)$ for
some $(a_0, a_1) \geq \mathbf{0}$. Due to the shape of
$\mathcal{R}_s$, this will result in one of the two UEs being
scheduled in each slot.

The rate region, time averaged over all channel states, $\overline{\mathcal{R}}$,
is shown in the right panel of
Figure~\ref{fig:two-ues-slot-and-time-average-rate-regions}, along
with the proportional fair operating point (for the utility
$U_i(r_i) = \ln(1 + r_i)$). The outer boundary of $\overline{\mathcal{R}}$ has
been obtained by simulating the above wireless channel model for
$10^{6}$~slots, and obtaining the operating points for the utility
$w_0 r_0 + w_1 r_1$ for a range of values of $(w_0, w_1)$.

\subsection{Two UEs:  Index-bias smoothing, \&  slow time scale updates}
\label{sec:two-ues-comparison-pf-rg-lm}

\begin{figure}[!h]
\begin{center}
\includegraphics[scale=0.07]{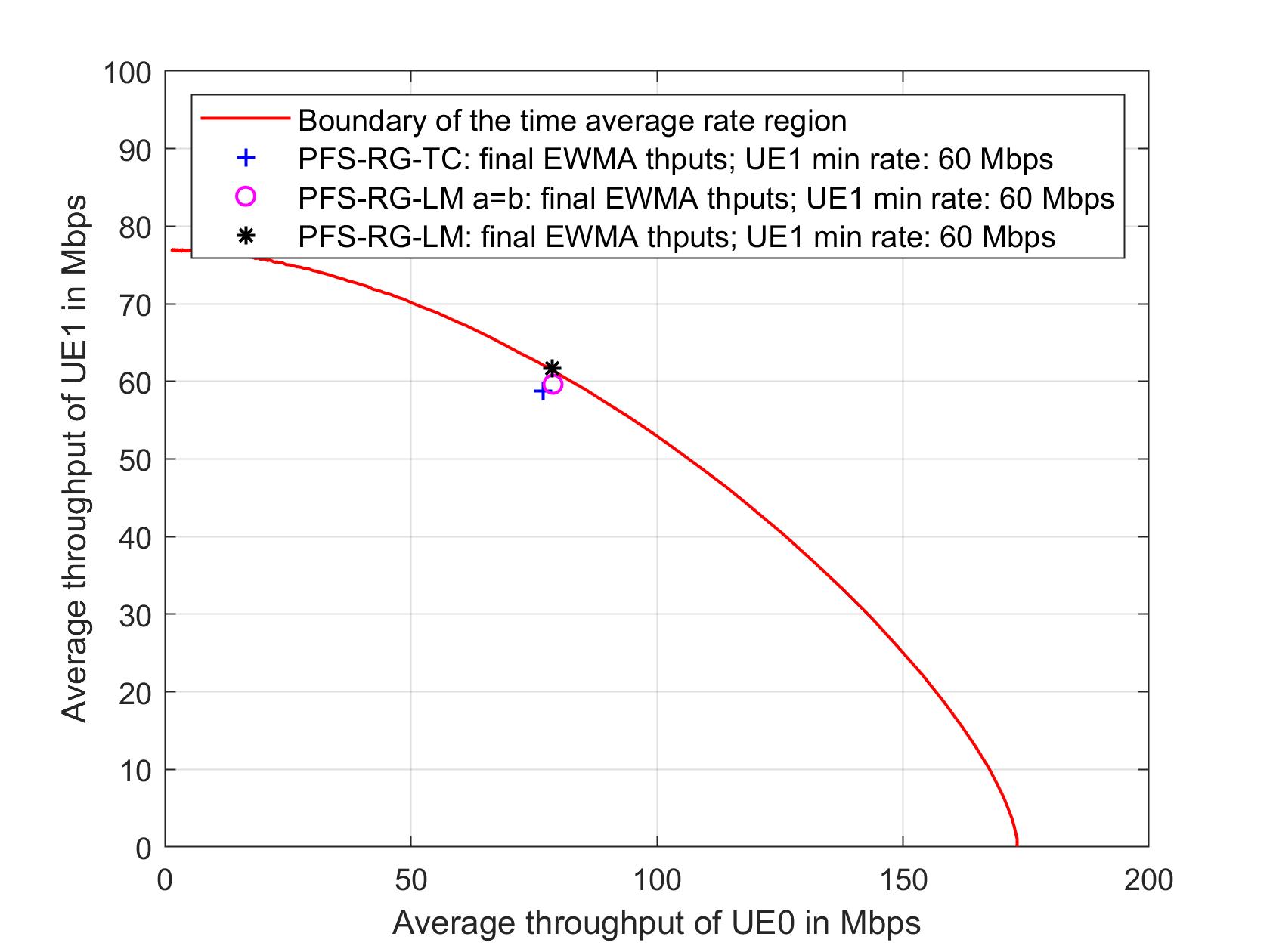}
\includegraphics[scale=0.07]{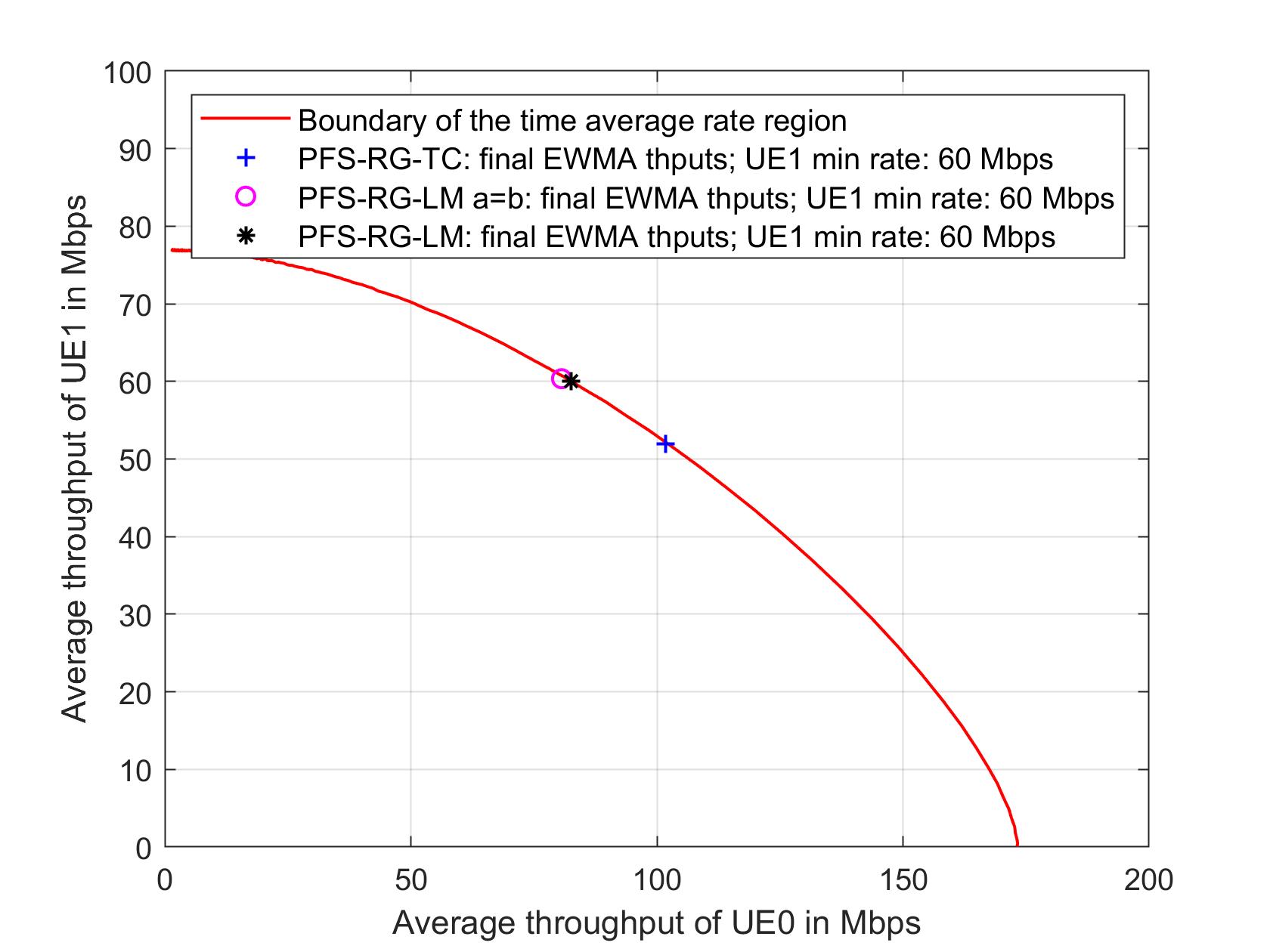}
\end{center}
% \begin{center}
% \includegraphics[scale=0.09]{../figures/rate-region_two-ues_rayleigh-fading_pf-rg-lm_and_pf-rg-tc_operating-points}
% %\hspace{1mm}
% \includegraphics[scale=0.09]{../figures/time-series_two-ues_rayleigh-fading_100mw_c0-42_eta-3_w-40_d-100-200_slot-001_pf-rg-lm_and_pf-rg-tc_a-0005_b-000005}
% %\hspace{1mm}
% \includegraphics[scale=0.09]{../figures/time-series_two-ues_rayleigh-fading_100mw_c0-42_eta-3_w-40_d-100-200_slot-001_pf-rg-lm_and_pf-rg-tc_a-0005_b-0000005}
% \end{center}
\caption{PF-RG-TC, PF-RG-LM with $a=b$, and PF-RG-LM: Left plots
  $a = 0.0005, b = 0.000005$; Right plots
  $a = 0.00005, b = 0.0000005$. $\text{UE}_1$ rate guarantee
  $60$~Mbps.  The final rate pairs provided by the three algorithms
  superimposed on $\overline{\mathcal{R}}$.}
\label{fig:pf-rg-tc_pf-rg-lm_operating-points}
\end{figure}

\begin{figure}[!h]
\begin{center}
\includegraphics[scale=0.07]{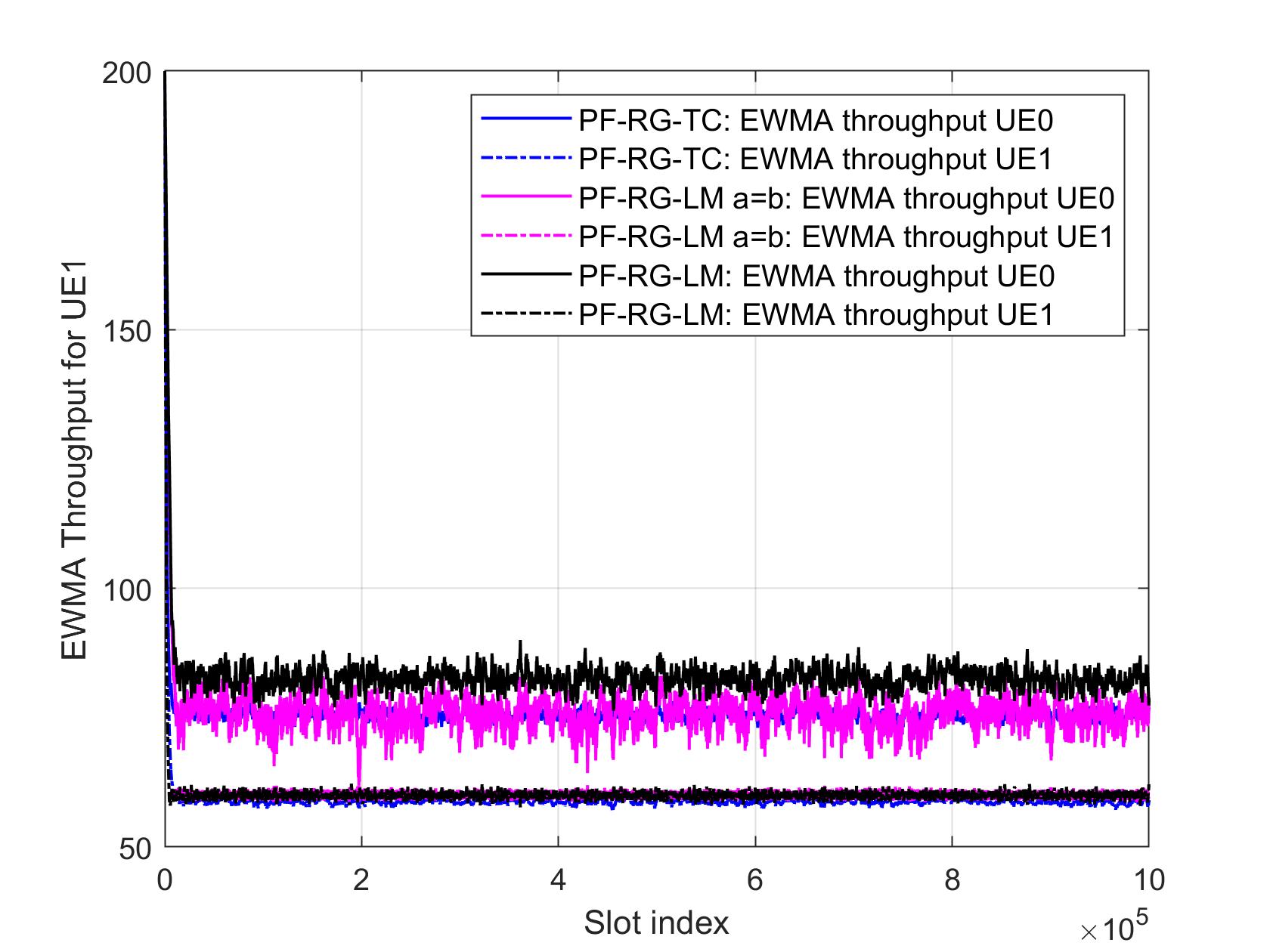}
\includegraphics[scale=0.07]{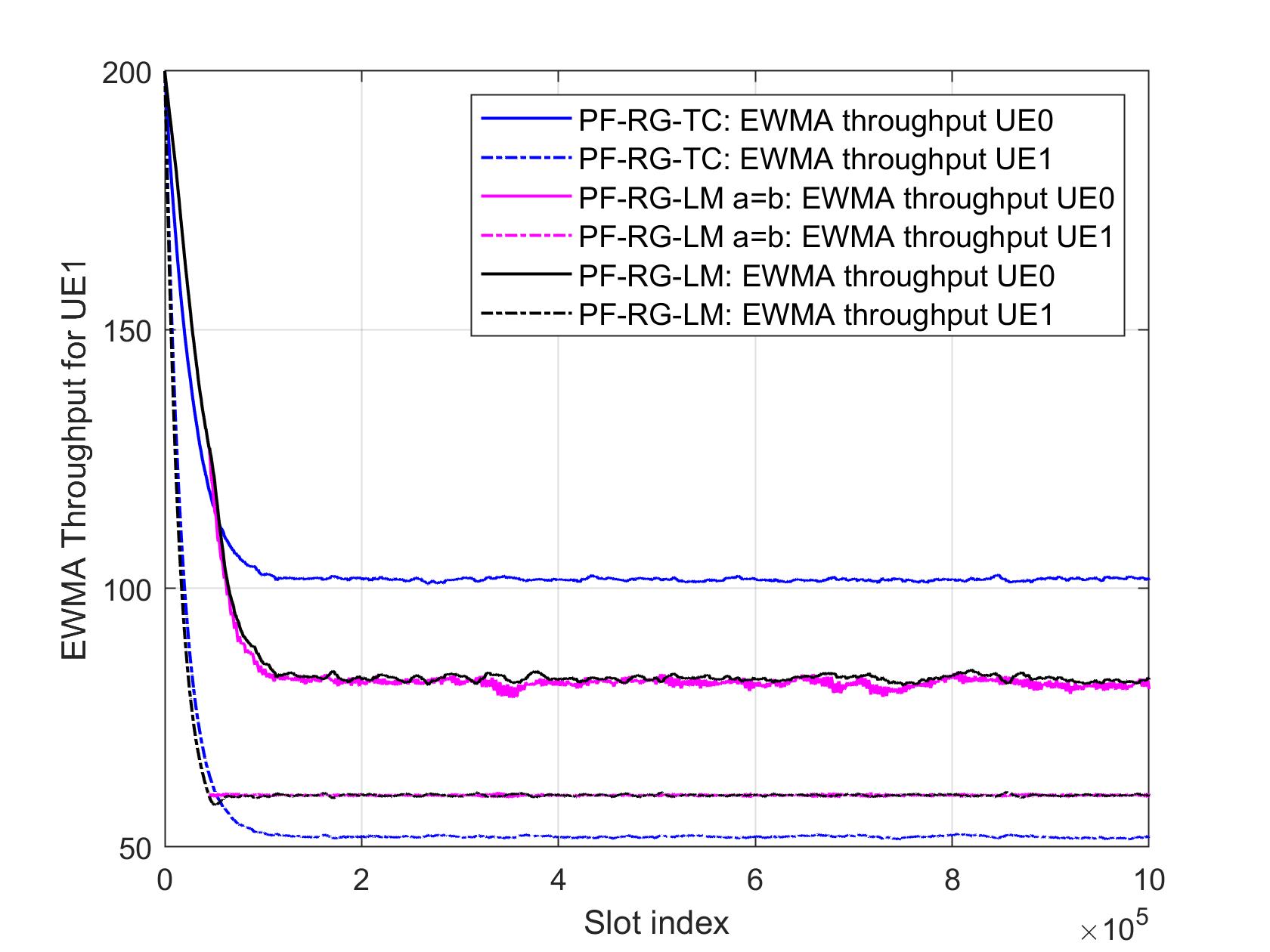}
\end{center}
\caption{PF-RG-TC, PF-RG-LM with $a=b$, and PF-RG-LM: Left plots
  $a = 0.0005, b = 0.000005$; Right plots
  $a = 0.00005, b = 0.0000005$. $\text{UE}_1$ rate guarantee
  $60$~Mbps.  Time series of EWMA throughputs (plots are in color). }
\label{fig:pf-rg-tc_pf-rg-lm_ewma-time-series}
\end{figure}

\begin{figure}[!h]
\begin{center}
\includegraphics[scale=0.07]{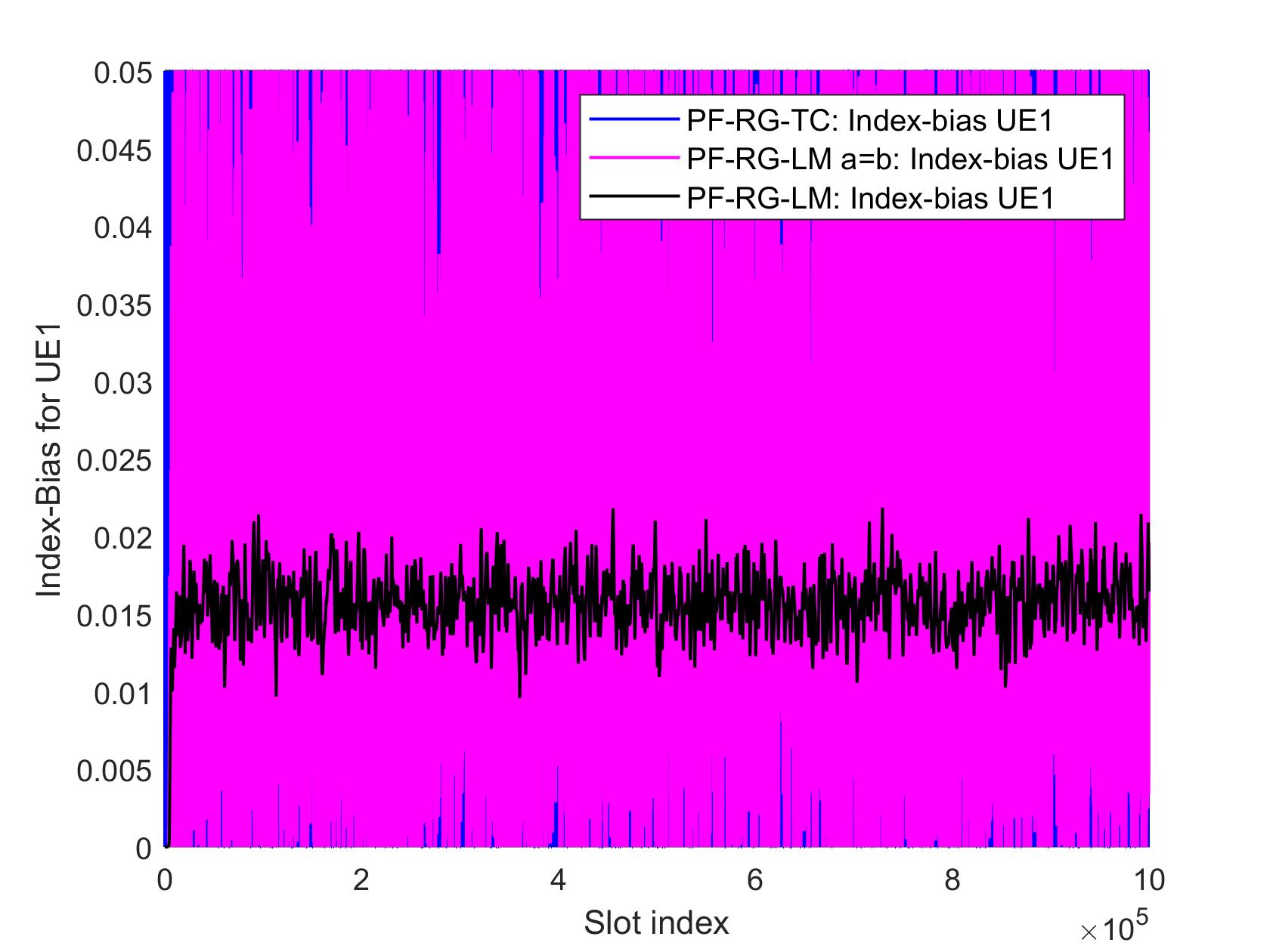}
\includegraphics[scale=0.07]{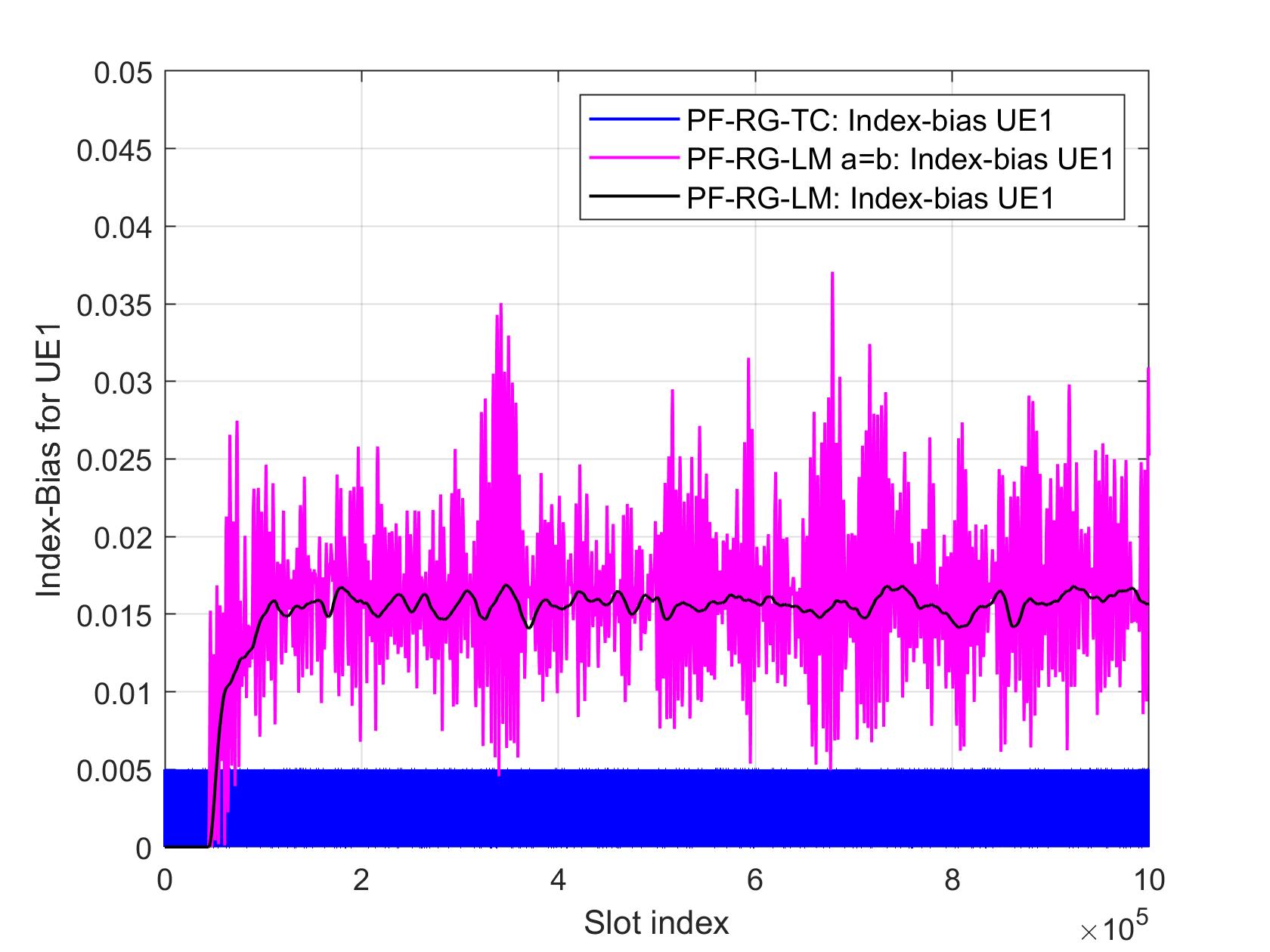}
\end{center}
\caption{PF-RG-TC, PF-RG-LM with $a=b$, and PF-RG-LM: Left plots
  $a = 0.0005, b = 0.000005$; Right plots
  $a = 0.00005, b = 0.0000005$. $\text{UE}_1$ rate guarantee
  $60$~Mbps.  Time series of index-biases (plots are in color).}
\label{fig:pf-rg-tc_pf-rg-lm_index-bias-time-series}
\end{figure}

In Figures~\ref{fig:pf-rg-tc_pf-rg-lm_operating-points},
\ref{fig:pf-rg-tc_pf-rg-lm_ewma-time-series}, and
\ref{fig:pf-rg-tc_pf-rg-lm_index-bias-time-series} we show the results
from running the PF-RG-LM (with $b \ll a$, and $ b = a$), and PF-RG-TC
algorithms on rate data generated by the wireless communication model
described in Section~\ref{sec:wireless-communication-model}, with the
BS transmit power set at $100$~mW. With $a = 0.0005, b = 0.000005$
(left plots) all three approaches provide an average rate of about
$60$~Mbps to $\text{UE}_1$, but PF-RG-TC and PF-RG-LM (with $a=b$)
provide lower rates to $\text{UE}_0$.
Figure~\ref{fig:pf-rg-tc_pf-rg-lm_index-bias-time-series} shows that
the index-bias for PF-RG-TC fluctuates wildly, not appearing to follow
any trend; with $a=b$, PF-RG-LM provides a slightly more controlled
but still highly fluctuating index-bias. With the index-bias being
updated at a much slower time-scale than the EWMA rates, the
index-bias for PF-RG-LM fluctuates slowly around a steady-state value.
With $a = 0.00005, b = 0.0000005$ (right plots) the throughputs
provided by PF-RG-TC are far from optimum.  PF-RG-LM with $a=b$ and
$b \ll a$ both provide accurate tracking of the optimum operating
point. The index-bias for PF-RG-LM with $a=b$ is better controlled but
still fluctuates significantly. On the other hand with a slower
time-scale update, PF-RG-LM not only provides an accurate tracking of
the optimum EWMA throughputs, but hovers around a value of
approximately $0.016$.

We may infer that the two-time scale stochastic approximation
algorithm used in PF-RG-LM provides not only an accurate throughput
operating point with rate guarantees, but also an accurate tracking
of the Lagrange multipliers. We notice, from
Equation~(\ref{eqn:pf-rg-tc-index-bias-update}), that in PF-RG-TC, in
addition to stochastic approximation not being used in the token
counter update, the token counter is updated by the rate allocation in
the previous slot, rather than the EWMA rate, as in PF-RG-LM (see
Equation~(\ref{eqn:index-bias-update})).

\subsection{PF-RG-LM with $N (> 2)$ UEs}
\label{sec:pf-rg-lm-n-UEs}

\begin{figure}[!h]
\begin{center}
  \includegraphics[scale=0.07]{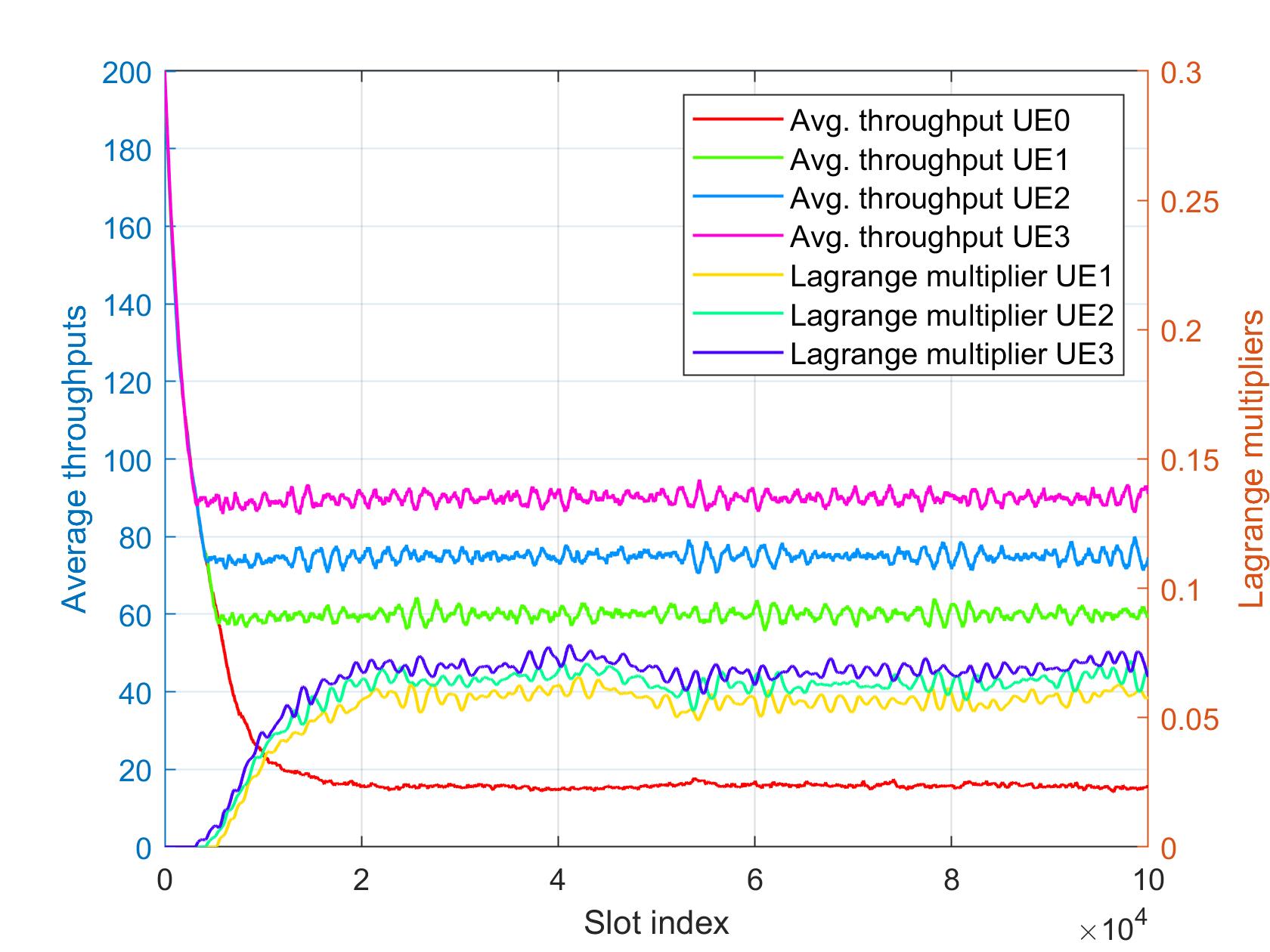}
  \hspace{1mm}
  \includegraphics[scale=0.07]{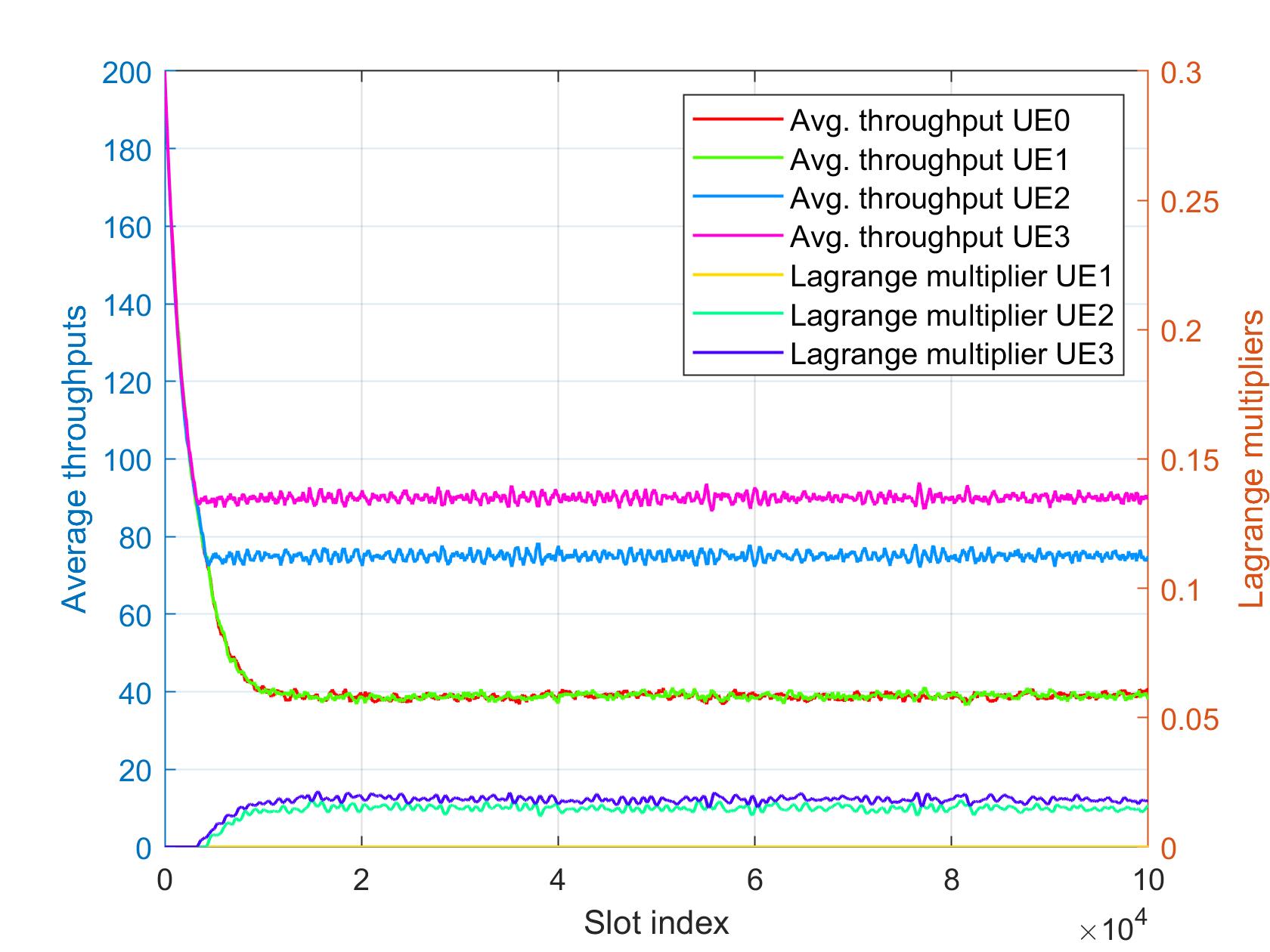}
\end{center}
\caption{PF-RG-LM with $a = 0.0005, b = 0.000005$, four UEs, at
  $200$~m, with rate guarantees $0, 60, 75, 90$~Mbps (left plot), and
  $0, 0, 75, 90$ (right plot). The plots (in color) show the time
  series of the EWMA throughputs of the four UEs, and
  $\nu_i(k), 2 \leq i \leq 4$. }
\label{fig:time-series_four-ues_0-60-75-90_rayleigh-fading_1000mw_c0-42_eta-3_w-40_d-200_slot-001_pf-rg-lm}
\end{figure}

In
Figure~\ref{fig:time-series_four-ues_0-60-75-90_rayleigh-fading_1000mw_c0-42_eta-3_w-40_d-200_slot-001_pf-rg-lm},
we provide results from running PF-RG-LM on rate data generated from
the wireless communication model described in
Section~\ref{sec:wireless-communication-model}, with 4 UEs, all at
$200$~m from the BS, and the BS transmit power set at $1000$~mW. In
the left plot, the rate guarantees are $0, 60, 75, 90$~Mbps for
$\text{UE}_0$, $\text{UE}_1$, $\text{UE}_2$, and $\text{UE}_3$, and in
the right plot these are $0, 0, 75, 90$. Both plots show that the
UEs with rate guarantees obtain their respective guaranteed rates. In
the left plot, $\text{UE}_0$ achieves a rate of a little over
$15$~Mbps, whereas for the case in the right plot, $\text{UE}_0$ and
$\text{UE}_1$ achieve the same rate of about $40$~Mbps. The Lagrange
multipliers also stabilize, with the larger guaranteed rate
corresponding to the larger Lagrange multipliers, as expected. Notice
that, in the right plot $\nu_1(k) = 0$ for all $k$.

\subsection{Polyhedral average rate region, $\overline{\mathcal{R}}$}
\label{sec:polyhedral-rate-regions}

\begin{figure}[!h]
\begin{center}
  \includegraphics[scale=0.07]{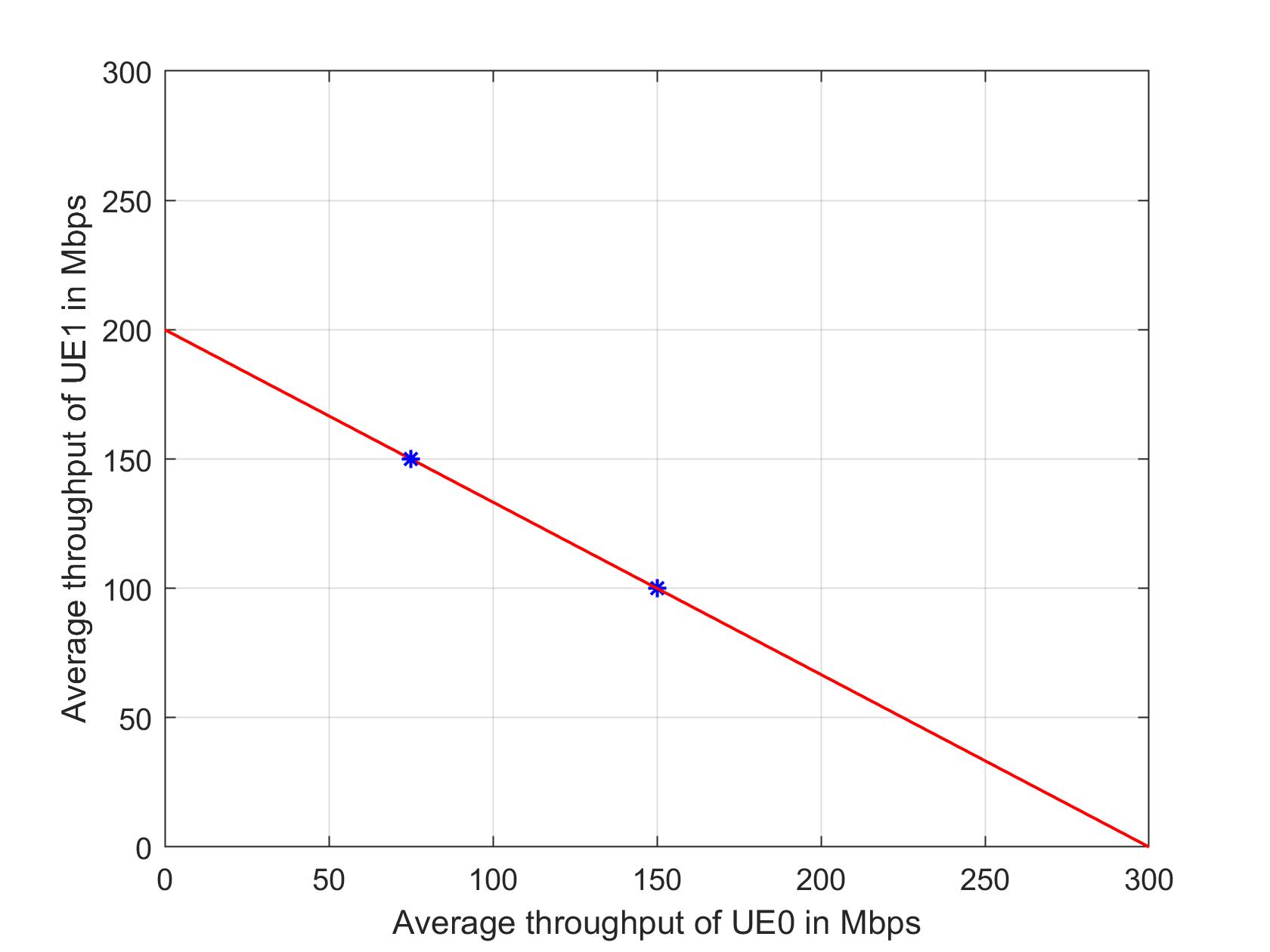}
  \hspace{1mm}
  \includegraphics[scale=0.07]{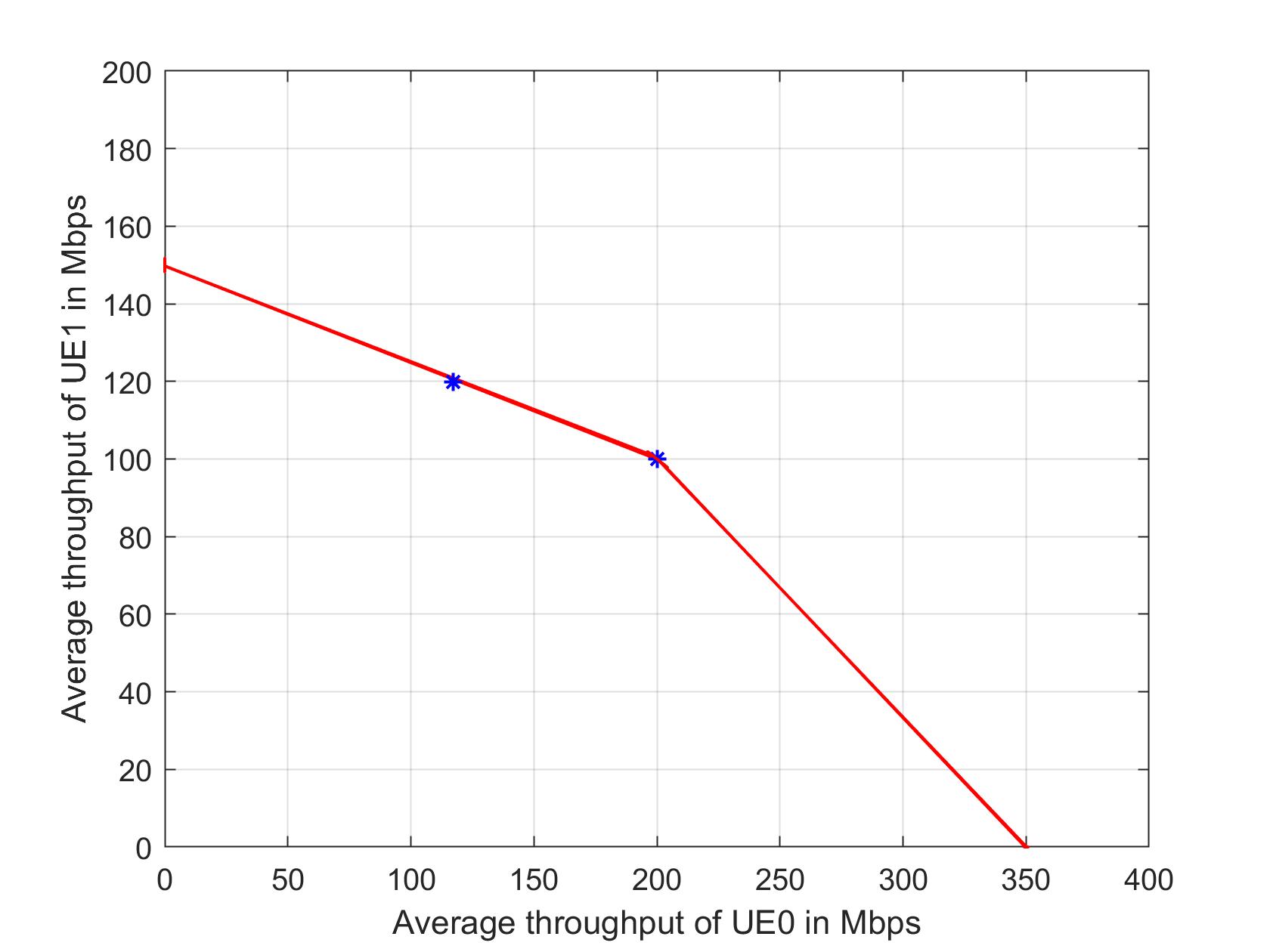}
\end{center}
\caption{Left plot: Two UEs, rate region with fixed rates $300$~Mbps
  and $200$~Mbps (i.e., the channel has only one state). Shown also
  are the PF throughputs, and the PF-RG-LM throughputs with RG of
  $150$~Mbps for $\text{UE}_1$. Right plot: Two UEs, rate region with
  two channel states ($(400, 100)$~Mbps, and $(300, 200)$~Mbps), with
  probabilities $0.5$.  Shown also are the PF throughputs, and the
  PF-LG-LM throughputs with RG of $120$~Mbps for $\text{UE}_1$.  We
  used $a = 0.0005, b = 0.000005$. }
\label{fig:two-ues_finite-channel-states}
\end{figure}

Figure~\ref{fig:two-ues_finite-channel-states} shows that the
algorithms with and without rate guarantees work correctly even when
the rate region, $\overline{\mathcal{R}}$, is not strictly convex. In
the left plot, the channel has one state, i.e.,
$\mathcal{S} = \{s'\}$, $\overline{\mathcal{R}} = \mathcal{R}_{s'}$
and, in each slot one of the UEs will be scheduled. However,
time-sharing occurs over time, yielding the desired throughputs. In
the right plot, the channel has two states, i.e.,
$\mathcal{S} = \{s', s''\}$; $\overline{\mathcal{R}}$ has the
ployhedral shape shown.  While the PF-RG-LM algorithm works correctly,
our convergence proof does not apply to this case. This remains an
item of our current work.

\section{Conclusion}
\label{sec:conclusion}

Assuming feasibility of the utility optimization problem, we have
proposed a slow time-scale update for the additive index-bias that
promotes rate guarantees in the usual gradient-type scheduling
algorithm. We provide a convergence proof of this algorithm for
strictly convex average rate regions. Our simulations show that the
slower time-scale index-bias update not only provides accurate maximum
utility throughputs, but the index-biases provide very good
approximations to the optimum Langrange multipliers. Practical
considerations might require that the throughput averaging time-scale
might be specified by the standard or by user requirements, in which
case, having a separate time-scale for the index-bias update would be
useful for providing a very good estimate of the optimum Lagrange
multiplier.

\section*{Appendix}

\subsection*{Global asymptotic stability}
\label{appsec:global-asymp-stability}

We now argue that, for a fixed $\boldsymbol{\nu}$, the
o.d.e.~\eqref{eqn:ewma-ode-with-fixed-index-bias} has a unique
globally asymptotically stable equilibrium.  The following proof has
been taken from \cite{winet.stolyar05gradient-scheduling}, and is
provided here for completeness. Recall
that~\eqref{eqn:ewma-ode-with-fixed-index-bias} is
  \[
    \dot{\boldsymbol{\theta}}(t) =
    \overline{\mathbf{h}}(\boldsymbol{\theta}(t), \boldsymbol{\nu}) =
    \arg \max_{\boldsymbol{r} \in \overline{\mathcal{R}}} [\nabla
    U(\boldsymbol{\theta}(t) + \boldsymbol{\nu}]^T \cdot \boldsymbol{r}
    - \boldsymbol{\theta}(t).
  \]
  Let $d(\boldsymbol{\theta}(t),\overline{\mathcal{R}})$ denote the distance of
  $\boldsymbol{\theta}(t)$ to the set $\overline{\mathcal{R}}$. We first argue
  that
  $d(\boldsymbol{\theta}(t), \overline{\mathcal{R}}) \leq
  d(\boldsymbol{\theta}(0), \overline{\mathcal{R}}) e^{-t}$.

  Let $\boldsymbol{\theta}_p(t)$ denote the projection of
  $\boldsymbol{\theta}(t)$ on $\overline{\mathcal{R}}$. Let
\[
  \boldsymbol{r}(t) = \arg \max_{\boldsymbol{r} \in \overline{\mathcal{R}}}
  [\nabla U(\boldsymbol{\theta}(t)) + \boldsymbol{\nu}]^T \cdot
  \boldsymbol{r}
\]
Observe that $\boldsymbol{\theta}(t)$ and $\boldsymbol{r}(t)$ are
continuous in $t$, the latter following from
Condition~\ref{cond:continuity-arg-max-over-rate-regions}. For
$0 \leq h \leq 1$,
$\boldsymbol{\rho}(t+h) := \boldsymbol{\theta}_p(t) + h
(\boldsymbol{r}(t) - \boldsymbol{\theta}_p(t)) \in
\overline{\mathcal{R}}$, and hence can be used to bound the distance
to $\overline{\mathcal{R}}$ at time $t+h$ for $h$ sufficiently
small. We can then upper bound the rate of change of the distance of
$\boldsymbol{\theta}(t)$ to the set $\overline{\mathcal{R}}$ as
follows:
\begin{eqnarray*}
  \lefteqn{\limsup_{h \downarrow 0} \frac{1}{h} \left[d(\boldsymbol{\theta}(t+h), \overline{\mathcal{R}}) - d(\boldsymbol{\theta}(t), \overline{\mathcal{R}}) \right]} \\
  & \leq & \limsup_{h \downarrow 0} \frac{1}{h} \left[\|\boldsymbol{\theta}(t+h) -  \boldsymbol{\theta}_p(t+h)\| - \|\boldsymbol{\theta}(t) - \boldsymbol{\theta}_p(t)\| \right] \\
  & = & -\| \boldsymbol{\theta}(t) - \boldsymbol{\theta}_p(t) \| \\
  & = & -d(\boldsymbol{\theta}(t), \overline{\mathcal{R}})
\end{eqnarray*}
where the penultimate equality requires an argument that uses the
continuity of $\boldsymbol{\theta}(t)$ and $\boldsymbol{r}(t)$ in $t$.

Thus, for any $\epsilon > 0$ and any bounded set $A$ on the positive
orthant, there exists a $T_1 = T_1(\epsilon, A)$ such that
$\boldsymbol{\theta}(t) \in \overline{\mathcal{R}}^{(\epsilon)}$ for
all $t \geq T_1$, where $\overline{\mathcal{R}}^{(\epsilon)}$ is the
$\epsilon$-thickening of $\overline{\mathcal{R}}$, which is the set of
all points that are within a distance $\epsilon$ of
$\overline{\mathcal{R}}$.

We recall Equation~\ref{eqn:ewma-ode-fixed-index-bias-stable-points},
which defines $\boldsymbol{\theta}_{\infty}(\boldsymbol{\nu})$ as the
maximizer of
$(U(\boldsymbol{\theta}) + \boldsymbol{\nu}^T \cdot
\boldsymbol{\theta})$ in $\overline{\mathcal{R}}$. Now, fix an
arbitrary $\delta > 0$, and denote by
$O_{\delta}(\boldsymbol{\theta}_{\infty}(\boldsymbol{\nu}))$ the open
ball of radius $\delta$ around the maximizer. Fix $\epsilon > 0$ small
enough so that the maximum of
$(U(\boldsymbol{\theta}) + \boldsymbol{\nu}^T \cdot
\boldsymbol{\theta})$ in
$(\overline{\mathcal{R}}^{(\epsilon)} \setminus
O_{\delta}(\boldsymbol{\theta}_{\infty}(\boldsymbol{\nu}))) \cap \mathbb{R}^M_+$, denoted
$h^*$, satisfies
$h^* < \max_{\theta \in \overline{\mathcal{R}}}
[U(\boldsymbol{\theta}) + \boldsymbol{\nu}^T \cdot
\boldsymbol{\theta}].$

Given that $\boldsymbol{\theta}(0) \in A$, a compact set, the dynamics
enters $\overline{\mathcal{R}}^{(\epsilon)}$ and remains there for all
$t > T_1(\epsilon, A)$. Further, given that so long as the dynamics
remains in
$(\overline{\mathcal{R}}^{(\epsilon)} \setminus
O_{\delta}(\boldsymbol{\theta}_{\infty}(\boldsymbol{\nu})) \cap
\mathbb{R}^M_+$, we have
\begin{eqnarray*}
  \lefteqn{\frac{d}{dt} [U(\boldsymbol{\theta}(t) + \boldsymbol{\nu}^T \cdot \boldsymbol{\theta}(t)]} \\
  & = & [\nabla U(\boldsymbol{\theta}(t)) + \boldsymbol{\nu}]^T \cdot [\boldsymbol{r}(t) - \boldsymbol{\theta}(t)] \\
  & \geq & [\nabla U(\boldsymbol{\theta}(t)) + \boldsymbol{\nu}]^T \cdot [\boldsymbol{\theta}_{\infty}(\boldsymbol{\nu}) -
           \boldsymbol{\theta}(t)] \\
  & \geq & (U(\boldsymbol{\theta}_{\infty}(\boldsymbol{\nu}))
           + \boldsymbol{\nu}^T \cdot \boldsymbol{\theta}_{\infty}(\boldsymbol{\nu})) -((U(\boldsymbol{\theta}(t)) +
           \boldsymbol{\nu}^T \cdot \boldsymbol{\theta}(t)) \\
  & \geq & (U(\boldsymbol{\theta}_{\infty}(\boldsymbol{\nu}))
           + \boldsymbol{\nu}^T \cdot \boldsymbol{\theta}_{\infty}(\boldsymbol{\nu})) - h^*  >  0,
\end{eqnarray*}
where the second inequality follows from concavity of
$(U(\boldsymbol{\theta}) + \boldsymbol{\nu}^T \cdot
\boldsymbol{\theta})$ in $\boldsymbol{\theta}$.  This implies that the
dynamics must exit
$(\overline{\mathcal{R}}^{(\epsilon)} \setminus
O_{\delta}(\boldsymbol{\theta}_{\infty}(\boldsymbol{\nu})) \cap
\mathbb{R}^M_+)$ at some time $t > T_1(\epsilon, A) + T_2$. Combining
this with the fact that the dynamics remains in
$\overline{\mathcal{R}}^{(\epsilon)}$ for all $t > T_1(\epsilon, A)$, we conclude
that
$\boldsymbol{\theta}(t) \in
O_{\delta}(\boldsymbol{\theta}_{\infty}(\boldsymbol{\nu}))$ for all
$t \geq T_1(\epsilon, A) + T_2$. Since $\delta > 0$ was arbitrary, we conclude that
$\boldsymbol{\theta}(t) \rightarrow
\boldsymbol{\theta}_{\infty}(\boldsymbol{\nu})$, which establishes the
desired global asymptotic stability.

\subsection*{Obtaining $\nu_{\text{max}}$: Two UE Example}

To illustrate how it can be determined if there is a bound on the
optimum Lagrange multipliers, we provide an illustration for two UEs,
$\text{UE}_0$ and $\text{UE}_1$ in which $\text{UE}_1$ has the minimum
rate guarantee $\theta_{1,\text{min}}$. By sensitivity analysis of the
optimum solution of the optimization problem
(\ref{eqn:optimization-problem}), we can conclude that:
\begin{eqnarray*}
  \frac{\partial}{\partial \theta_{1,\text{min}}}
  (U(\theta^*_0(\theta_{1,\text{min}})) + U(\theta^*_1(\theta_{1,\text{min}}))) &=& -\nu^*_1
\end{eqnarray*}
where
$\left(\theta^*_0(\theta_{1,\text{min}}),
  \theta^*_1(\theta_{1,\text{min}})\right)$ is the optimum primal
solution, and $\nu^*_1$ the  optimal dual solution. It follows that
\begin{eqnarray*}
  U^\prime(\theta^*_0(\theta_{1,\text{min}})) \frac{\partial}{\partial \theta_{1,\text{min}}}
  \theta^*_0(\theta_{1,\text{min}}) + U^\prime(\theta^*_1(\theta_{1,\text{min}})) \cdot 1 &=& -\nu^*_1
\end{eqnarray*}
Where, in the second term, we have assumed that
$\theta_{1,\text{min}}$ is strictly feasible, i.e., the rate guarantee
can be increased. At the point
$\left(\theta^*_0(\theta_{1,\text{min}}),
  \theta^*_1(\theta_{1,\text{min}})\right)$ on the boundary of the
average rate region $\overline{\mathcal{R}}$ let
$(w^*_0, w^*_1) > \mathbf{0}$ be the normal to the boundary. It
follows that
$\frac{\partial}{\partial \theta_{1,\text{min}}}
\theta^*_0(\theta_{1,\text{min}}) = - \frac{w^*_0}{w^*_1} < 0$, which yields
\begin{eqnarray*}
  \nu^*_1 \leq U^\prime(\theta^*_0(\theta_{1,\text{min}})) \frac{w^*_0}{w^*_1}
\end{eqnarray*}
One way to satisfy Condition~\ref{cond:nu-max-existence} is to ensure
that the right hand side of the above inequality is bounded.

%\newpage
%\pagebreak
\bibliographystyle{plain}
\bibliography{winet,books}

\end{document}